# Graphene-based Antennas for Terahertz Systems: A Review

Diego Correas-Serrano and J. Sebastian Gomez-Diaz

Department of Electrical and Computer Engineering,

University of California, Davis, Davis, CA 95616 USA

(Email: dcorreas@ucdavis.edu, jsgomez@ucdavis.edu)

*Abstract*— **We review the use of graphene to develop reconfigurable, miniaturized, and efficient terahertz (THz) antennas and associated feeding networks, and attempt to identify current research trends and mid- and long-term challenges and prospects. We first discuss the state of the art in resonant, leaky-wave and reflectarray antennas, providing a critical assessment of their performance, limitations, and main challenges that remain to be addressed. Next, we examine different integrated feeding networks, including components such as switches, filters, and phase shifters, and we clarify the impact that graphene's intrinsic spatial dispersion may have in their performance. Our outlook clearly describes how graphene can bring exotic functionalities to all these devices, including quasi real-time reconfiguration capabilities and magnet-less non-reciprocal responses. Some exciting applications of THz antennas are then presented and discussed, including transceivers, biosensors, and first experimental realizations of detectors and modulators. We conclude by outlining our vision for the promising future of graphene-based THz antennas.**

## I. INTRODUCTION

Advanced electromagnetic antennas operating at micro- and millimeter wave frequencies have become indispensable in today's information society [1], [2]. They are massively employed in a wide range of applications including multimedia broadcasting, mobile or satellite communications, radars, environmental sensing, or medical systems [3]–[14]. This plethora of applications has naturally triggered enormous research and technological progress in the field in the past decades. However, the exponential growth of the information society is continuously imposing very challenging – and sometimes conflicting – technological demands, such as higher communication data rates, higher number of functionalities in a single, miniaturized device, future 5G heterogeneous networks, fully integrated and miniaturized sensing solutions, and the efficient use of the spectrum and energy resources, to name a few.

One straightforward solution to fulfill such demands would be to further increase the operation frequency of current antennas, taking advantage of the terahertz (THz) spectrum [15]–

[18], i.e., electromagnetic (EM) waves with frequencies ranging from 0.1 to 10 THz [16], [19]. In fact, this band presents unique opportunities for advanced applications [16], [19], including wireless communications with hundreds of gigabit per second or even terabit-per-second data rates, high-speed miniaturized processing systems, sensing the presence of specific chemical or biological processes, ultra-high resolution imaging systems for inspection, screening, and non-invasive identification of materials in real-time by spectroscopy, study of cosmic background radiation, or detecting gases in the Earth and in the atmosphere of other planets, among many others [16]–[22]. Despite recent promising advances [21], these applications are not currently being fully-exploited by the information society due to the immature state of THz technology in terms of antennas, detectors, sources, and basic components. In fact, this frequency region is known as the "THz gap", as it occupies a technology gap between the well-developed areas of electronics and photonics.

In a related context, graphene, a two-dimensional carbon material, has recently triggered very intense and multidisciplinary research efforts due to its outstanding electromagnetic, mechanical, electrical, and thermal properties [23]–[32]. Graphene supports the propagation of surface plasmon polaritons (SPPs) able to exhibit strong wave localization, moderate loss, and the exceptional property of being tunable through electrical/magnetic bias or chemical doping. Importantly, since such plasmonic response appears in the terahertz and infrared frequency bands, graphene has quickly positioned itself as a very promising platform for THz transceivers and optoelectronic systems. Even though SPPs can be supported by composite structures at THz [33], graphene is yet the only pristine material able to provide such plasmonic response, thus opening exciting and unexpected prospects for wave manipulation and radiation in this frequency band. Graphene's unique properties have been exploited to propose miniaturized and reconfigurable resonant [34]–[40], leaky-wave [41]–[46], and reflectarray antennas [47]–[50] with unrivaled radiation efficiency and functionalities at THz. These antennas resonate at frequencies much lower than their metallic counterparts and they fully exploit the reconfigurable nature of graphene's conductivity [51]. In addition, all fundamental building blocks of front-end transceiver architectures can be realized using graphene plasmonics, including switches, phase-shifters, low/band pass filters, or modulators. Non-reciprocal propagation and giant faraday rotation have also been demonstrated when graphene is biased by magnetostatic fields [52]–[54], but the bulky magnets required with this approach are antithetical to the desirable miniaturization of graphene plasmonics. Even more excitingly, graphene's unique properties have led to the development of magnet-free non-reciprocal guided devices and antennas based on the spatiotemporal modulation of its conductivity, allowing to break the fundamental principle of time-reversal symmetry (reciprocity) without ferromagnetic materials or magneto-optical effects [43], [46], [55]–[62]. The adequate integration of all these components and functionalities into a single, miniaturized, device will significantly push the boundaries of current terahertz technology, paving the way towards silicon-compatible THz communication and sensing systems with unprecedented performance and responses.

It is important to mention that, to date, most graphene-based antennas, devices, and detectors have only been proposed, designed, and analyzed theoretically, and their electromagnetic response have been predicted using numerical simulations. However, we do highlight that recent experiments have fully verified the propagation of confined and tunable SPPs in graphene [63]–[66], which is the physical phenomenon that all antennas and components described here rely on. Earlier experiments yielded graphene of very poor quality that was not viable for any electrodynamic application other than to serve as a tunable resistor, severely limiting the practical feasibility of most research. The past few years, however, have seen immense progress in this regard, thanks to extensive efforts by dozens of groups worldwide to improve fabrication techniques. For instance, mobility improvements of several orders of magnitude have been reported for graphene structures sandwiched between hexagonal boron nitride layer [63], [67]–[69], which are well above the quality requirements of all antennas and devices discussed in this review. More recently, and despite the challenging fabrication and measurement techniques involved, the response of some exciting THz detectors [70], [71] and modulators [72]–[78] have already been experimentally confirmed, clearly outperforming previous state of the art and showing excellent agreement with numerical expectations. These pioneering experimental works have just barely unveiled part of the immense potential of THz plasmonics using graphene or even other two-dimensional materials such as black phosphorus [79]–[83] or 2D chalcogenides and oxides [84]–[89], confirming that this exciting field is still in its infancy and that the future of graphene and 2D materials in antenna research looks brighter than many ever dared to imagine during its early steps

In the following, we summarize and contextualize the evolution of THz graphene-based antenna research since its beginnings merely 6 years ago, and attempt to identify current research trends and mid- and long-term challenges and prospects. We review the state of the art in resonant antennas, leaky wave antennas and reflectarrays, and conclude with an overview of their feeding networks and the most exciting applications envisioned for graphene antennas and THz systems.

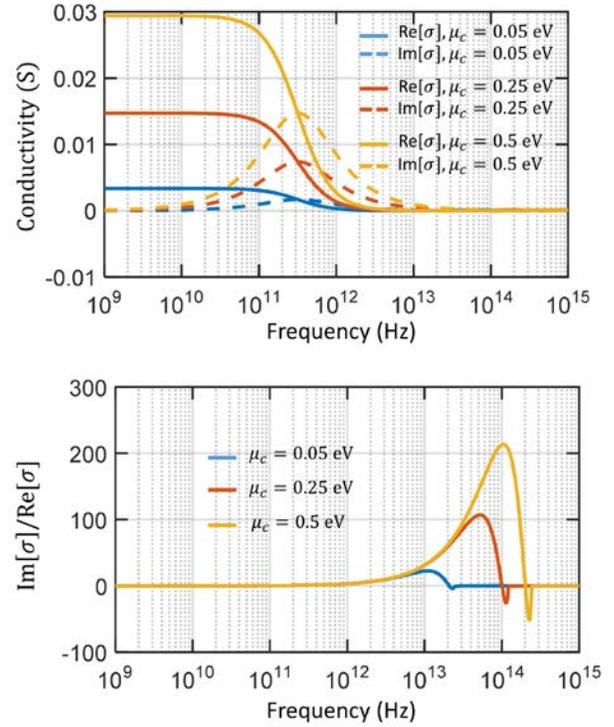

Fig. 1. Complex conductivity of graphene evaluated using the Kubo formula for several chemical potentials $\mu_c$, and ratio between its real and imaginary parts. Graphene relaxation time is $\tau = 0.5$ ps.

## II. ABOUT GRAPHENE CONDUCTIVITY

Graphene's conductivity directly depends on its unique band structure [26], [90]–[92] and a set of parameters such as chemical doping, Fermi energy (chemical potential), electron mobility, relaxation time, operation frequency or the environmental temperature. We stress that these parameters are not independent, but might be related to each other [26], [30], [90], [93], [94]. Using a semiclassical, local model and in the absence of magnetic bias, graphene's conductivity can be modeled at THz frequencies just using the intraband term of the Kubo formula [25], [95]

$$\sigma = -j\frac{e^2 k_B T}{\pi \hbar^2(\omega - j\tau^{-1})}\ln\left\{2\left[1+\cosh\left(\frac{\mu_c}{k_B T}\right)\right]\right\}, \quad (1)$$

where $e$ is the electron charge, $\tau$ is the electron relaxation time in graphene, $k_B$ is Boltzmann's constant, $T$ is temperature, $\hbar$ is the reduced Planck's constant, $\omega$ is the angular frequency, and $\mu_c$ is graphene's chemical potential. This expression is valid if $\hbar\omega < 2|\mu_c|$, which holds at THz frequencies for moderate $\mu_c$. At higher frequencies, interband contributions of graphene conductivity becomes significant and must be considered [26]. An example of graphene's dispersive conductivity is shown in Fig. 1. An exhaustive overview of such conductivity and associated phenomena is out of the scope of this paper. Instead, we refer the reader to some of the excellent literature available on this topic [25], [91], [92], [95], [96]. For the sake of completeness, we will briefly elaborate on some of its exciting properties that have direct application in antenna research.

From an electromagnetic perspective, graphene is frequency-dependent and behaves as a surface resistor at micro and millimeter wave frequencies. However, its conductivity acquires a different response at terahertz and infrared frequencies because its intrinsic kinetic inductance (associated to the imaginary part of the conductivity) plays the role of negative real permittivity in a bulk material. Consequently, and similarly to noble metals at optics, graphene supports the propagation of SPPs at these frequencies, which has recently triggered the development of antennas and guided devices. Specifically, in a free space standing graphene sheet of conductivity $\sigma$, the dispersion relation and characteristic impedance of the supported transverse magnetic (TM) plasmons can be computed as [25], [96], [97]

$$k_{SPP} = k_0 \sqrt{1 - \left(\frac{2}{\eta\sigma}\right)^2} \qquad (2)$$

$$Z_C = \frac{k_{SPP}}{\omega\varepsilon_0\varepsilon_{eff}} \qquad (3)$$

where $k_0$ and $\eta$ are the free-space wavenumber and impedance, $\varepsilon_0$ is the permittivity of vacuum, and $\varepsilon_{eff}$ is the effective permittivity of the surrounding media. Even though this simple model assumes laterally infinite sheets, it is nonetheless an extremely useful estimate of the dominant plasmon wavelength and loss in realistic structures like nanoribbons and patches.

Undoubtedly, one of the most exciting properties of graphene is the possibility to manipulate its complex conductivity. This can easily be done exploiting graphene's field effect by simply applying an external electrostatic field bias perpendicular to the sample [98]. The bias injects electrons or holes on the sheet, thus modifying graphene's chemical potential (see Fig. 1). Recent experiments have confirmed that this tunability can be very significant and fast – up hundreds of gigahertz [73], [99] –, thus enabling reconfiguration capabilities in all graphene-based antennas and components. In addition, when a magnetostatic bias is applied to graphene, its band structure is not continuous anymore but discretized in the so-called Landau levels [100]–[109]. Graphene's conductivity becomes then a gyrotropic tensor and leads to exciting non-reciprocal responses such as Faraday rotation or isolators [110], [111]. Optical pumping has also been used to modify graphene's properties, enabling THz lasing and operation as reconfigurable gain media [112]–[115]. It is also worth mentioning the non-local response of graphene [96], [116]–[118], which becomes important when the supported ultra-confined SPPs presents variations at speeds comparable to the carriers on the materials. This phenomenon must be considered when designing guided devices, as described later, since it can significantly alter their expected response.

## III. RESONANT ANTENNAS

Resonant antennas are ubiquitous in daily life, so it is only natural that they were the first kind of graphene antennas theoretically investigated. Pioneering works by Jornet and co-workers demonstrated that subwavelength graphene patches can resonate at THz frequencies, making them very attractive antennas for this elusive frequency band [34], [35], [119]. In their initial studies about graphene nano-ribbon (GNR)

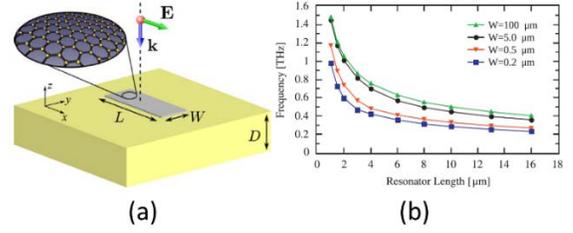

Fig. 2. Graphene-based patch as a THz scatterer [34]. (a) 3D schematic of the structure [39]. (b) Resonant frequency of the patch versus its geometrical features [39]. Graphene chemical potential and relaxation time are 0.0 eV and 0.1 ps, respectively. Reprinted from [39] with permission from Elsevier.

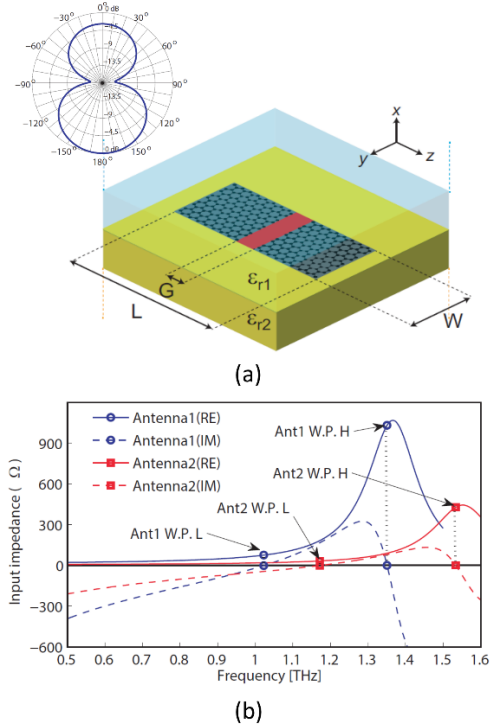

Fig. 3. Single layer graphene-based planar dipole [37]. (a) 3D schematic of the structure. A THz photomixer – red region with width $G$ – is employed to excite the antenna. Inset shows the E-plane radiation pattern. (b) Input impedance of the graphene dipoles described in Table 1. © Reprinted from [40], with the permission of AIP publishing.

dipoles, they did not yet consider the direct detection/excitation of SPPs for operation in reception/transmission, studying instead the extinction cross section of the structure to determine its resonant frequency. The configuration they considered is depicted in Fig. 2, and is comprised of a graphene ribbon transferred onto a substrate of thickness $D$. Fig. 2b shows the resonant frequency versus the ribbon width $W$ and length $L$. The ability of graphene patches and strips to resonate in the THz band arises from the unusually high kinetic inductance of graphene, which enables short-wavelength SPPs in this frequency range [120]–[122]. The associated resonant frequencies can easily be approximated by considering the propagation of SPPs along patches of finite length and then computing their Fabry-Perot resonances [39]. Examples of similar analysis using common transmission line techniques will be further described below. Once these resonant frequencies are available, the effective

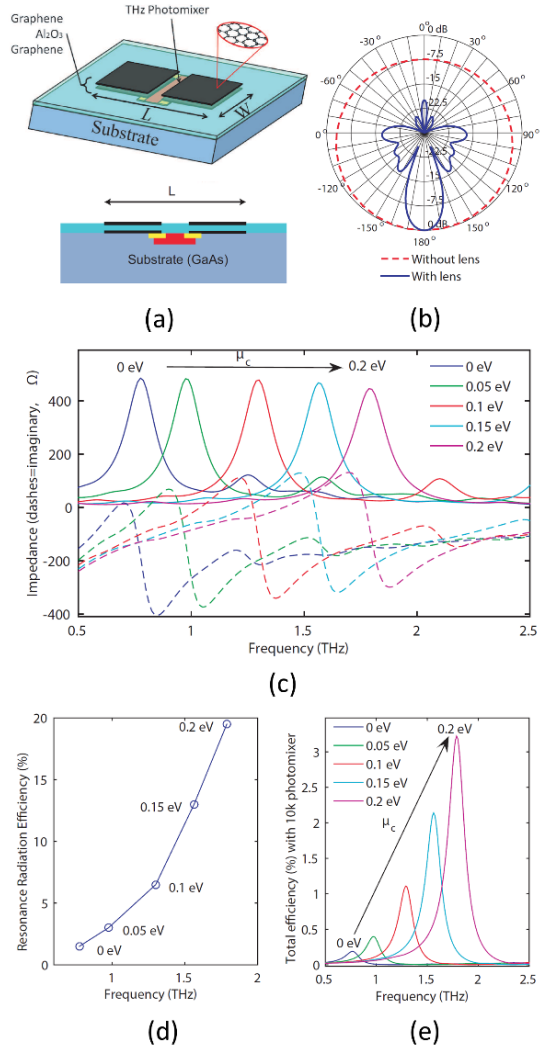

**(a)** **(b)**

**(c)**

**(d)** **(e)**

Fig. 4. Double layer graphene-based planar dipole [38] . (a) 3D schematic (top) and lateral view (bottom) of the structure. (b) E-plane radiation pattern with (solid blue line) and without (dashed red line) including a silicon lens to the antenna. The lens has a height of 572 $\mu m$, a radius of 547 $\mu m$ and is made of Silicon modelled with $\varepsilon_r = 11.66$ and $\tan\delta = 0.001$. The dielectric thickness is $t=160$ $\mu m$. (c) Input impedance of the graphene antenna upon variation of the chemical potential. Radiation (d) and total (e) efficiency – considering a 10 k$\Omega$ photomixer as a source – at resonance versus chemical potential. The antenna has a total length $L=11$ $\mu m$ and a width $W=7$ $\mu m$ and is composed of two stacked graphene patches [76] separated by a $h=100$ nm thick $Al_2O_3$ layer ($\varepsilon_r = 9$ and $\tan\delta = 0.01$). © Reprinted from [38], with the permission of AIP publishing.

| $\mu_c(eV)$ | L($\mu m$) | W($\mu m$) | W.P. | f(THz) | L/$\lambda_0$ |
|---|---|---|---|---|---|
| 0.13 | 17 | 10 | L | 1.023 | 0.06 |
| | | | H | 1.35 | 0.08 |
| 0.25 | 25 | 20 | L | 1.172 | 0.09 |
| | | | H | 1.534 | 0.12 |

Table 1. Characteristics of the single layer graphene-based planar dipoles studied in Fig. 3.

polarizability of the subwavelength patch can be retrieved, thus allowing to easily obtain the absorption and scattering cross sections of the structure.

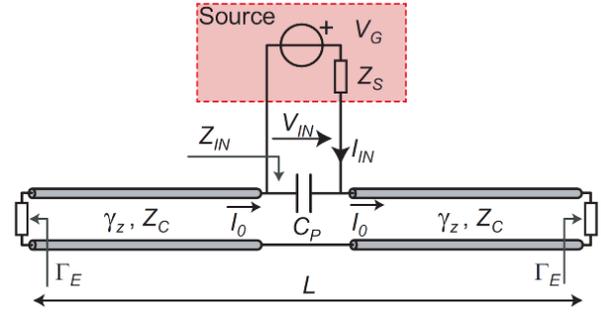

Fig. 5. Circuit model of graphene-based plasmonic dipoles [124]. © 2014 IEEE. Reprinted, with permission, from [124].

Soon after this study, Tamagnone and colleagues proposed the first true graphene antenna, in the sense that it bridges the gap between SPPs propagating on the sheets and far field radiation [37]. The geometry considered, depicted in Fig. 3, is similar to Fig. 2 and is based on the same operation principle. The key contribution is the addition of a port able to efficiently excite SPPs over graphene (or detect them in reception, from reciprocity). This allowed to perform the first quantitative study of input impedance and radiation efficiency of GNR dipole antennas, demonstrating unprecedented tunability via electrostatic bias and high values of input impedance suitable for THz sources such as photomixers. As an example, the input impedance of a graphene dipole antenna tuned to $\mu_c = 0.13$ and 0.25 eV is shown in Fig. 3b. The frequency points suitable for operation as an antenna (i.e., real input impedance) are marked in the plot as WPL and WPH (working points low and high, respectively), corresponding to the conditions $L \approx 0.5\lambda_{SPP}$, and $L \approx \lambda_{SPP}$ where $\lambda_{SPP}$ is the wavelength of the SPP mode supported by the graphene strip. As shown in Table 1, the dimensions of this antenna $\left(\frac{L}{\lambda_0}\right)$ remain deeply subwavelength. These promising findings encouraged the systematic study of graphene at THz from the unexplored perspective of antenna engineering.

This graphene patch antenna was refined shortly after by considering a double layer graphene stack and including a dielectric lens to increase its directivity [38] (see Fig. 4). This way, the advantages of wide tunability and real input impedance in subwavelength plasmonic devices could also be exploited in applications requiring directive antennas. Moreover, they demonstrate excellent impedance stability under wide reconfiguration of the resonant frequency, as shown in Fig. 4c. A more comprehensive study on the radiation and total efficiency of the proposed antenna as a function of chemical potential is summarized in Fig. 4d-e. We do highlight that the computed radiation efficiency is indeed high, clearly outperforming other antenna of similar size operating at THz [123]. Besides, it is interesting to note that the efficiency, unlike the input impedance, is not stable upon reconfiguration. This is easily explained in terms of electrical size of the antenna. Increasing the chemical potential $\mu_c$ increases the plasmon wavelength, upshifting the resonant frequency and increasing the electrical size of the antenna in terms of free-space wavelength. This is a common theme in all graphene antennas: extremely miniaturized devices can be designed by exploiting the high inductance of graphene for small $\mu_c$, but it usually comes at the cost of radiation efficiency. A rarely mentioned downside of using large $\mu_c$ is that the properties of graphene SPPs become less sensitive to bias voltages, negatively affecting electronic reconfigurability.

The design and analysis of graphene-based dipoles discussed so far relied heavily on full-wave simulations to compute the input impedance and the resonant frequency. In 2014, a simple and elegant circuit model for this type of antenna was proposed [124]. This model, depicted in Fig. 5, introduces a capacitance between the two dipole arms and a reactance at their termination to capture the effect of fringing fields. The characteristic of the transmission line – which represents SPPs propagation through the graphene patch – can be calculated exactly from numerical eigensolvers, or approximately by Eqs. (1)-(3) for the case of laterally infinite sheets. With these elements, the antenna resonant frequency, input impedance, and radiation/total efficiency can be calculated using straightforward transmission line theory.

To further increase the radiation efficiency of these antennas, the concept of graphene coated wires (GCW, see Fig. 6) antennas was recently introduced [40]. Authors analytically developed the dispersion relation of multiple variations of this cylindrical waveguide configuration supporting electrical and magnetic biasing, and showed that the absence of edges substantially reduces the propagation loss of SPPs compared to strip-based configurations [40], [125]–[128]. Moreover, this class of waveguides can be designed to be immune to the dielectric loss of the wire material, since all power travels outside [125]. These findings have important implications in the design of optimal THz interconnects and reconfigurable waveguide devices, but also in the design of resonant antennas. While the patch-based antennas have the inherent advantage of planarity, GCW dipole antennas were shown to present significantly higher radiation efficiency than similar planar implementations. Table 2 presents a performance summary for two comparable planar and GCW resonant antennas. GCWs presents significantly better performance, attributed to the their higher SPPs quality factor and increased robustness to dielectric loss and possible trapping of radiation by the substrate. Fig. 6c-d depict the input impedance of similar GCW- and patch-based dipoles, demonstrating an extended reconfiguration range for the GCW. Fig. 6e-f show the input impedance and radiation resistance [1] computed for a wide range of frequencies and $\mu_c$. The maxima and minima of radiation resistance in Fig. 6f roughly correspond to the odd and even multiples of the plasmon wavelength along the dipole, respectively. Like all resonant dipoles based on deeply confined plasmons, the radiation nulls appear because the contributions of adjacent subwavelength current units tend to cancel out in the far-field, which is not the case in conventional dipole antennas [40], [129], [130]. We note that a circuit model similar to Fig. 5 was used to design and characterize this class of resonant antennas (thin traces in Fig. 6c), further demonstrating the general validity of traditional microwave concepts in modeling graphene antennas.

We have presented an overview of the most fundamental graphene-based resonant antennas, but many groups worldwide have pursued ways to improve their performance or to achieve exotic responses. The variations are too numerous to describe in detail here [131]–[138], but include such interesting concepts as enhancing the radiation of quantum systems [139], an efficient integration of a photomixer in the graphene antenna [140], or non-reciprocal behavior based of magnetic bias with well-defined fundamental bounds [141], [142].

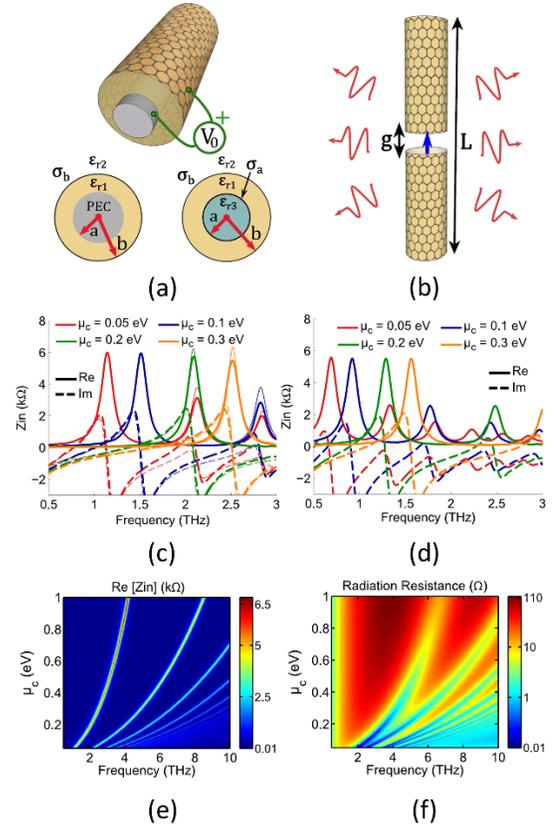

Fig. 6. Reconfigurable plasmonic antennas based on graphene tubes [40]. (a) 3D schematic of a graphene-based cylindrical waveguide. Insets illustrate the lateral view of a single (left) and dual (right) layer graphene cylindrical waveguides. (b) Operation of the device as an antenna. (c) Input impedance of a double-layer graphene cylindrical dipole upon reconfiguration. Thick traces computed with CST MWS, and thin traces computed with an equivalent circuit (see Fig. 5 and [40]). Parameters are $L$=26 $\mu m$, $a$=450 nm, $b$=500, $\varepsilon_{r1} = \varepsilon_{r3} = 3.8$. (d) Input impedance of a similar-sized double-plane graphene planar antenna (see Fig. 3 and [38]). Parameters are $W$=1 $\mu m$, permittivity and thickness of substrate $\varepsilon_r = 3.8$ and $t$=1 $\mu m$, and a distance between graphene layers $h$=50 nm. Real part of the input impedance (e) and radiation resistance (b) of the plasmonic dipole defined in panel (c) versus frequency and chemical potential. © 2015 IEEE. Reprinted with permission, from [40].

| $\mu_c$(eV) | GCW | | | Strip | | |
|---|---|---|---|---|---|---|
| | $V_0$(V) | $\eta_{rad}$ | $\eta_{tot}$ | $V_0$(V) | $\eta_{rad}$ | $\eta_{tot}$ |
| 0.05 | 0.46 | 1.2 | 1.1 | 0.43 | 0.16 | 0.12 |
| 0.1 | 1.84 | 4 | 3.7 | 1.75 | 0.5 | 0.4 |
| 0.2 | 7.3 | 14 | 12.5 | 7.0 | 2 | 1.4 |
| 0.3 | 16.5 | 26 | 24 | 15.7 | 3.8 | 3.5 |

Table 2. Bias voltages ($V_0$ in Fig. 6a) and efficiencies (%) of the graphene cylinder- and strip-based antennas described in Fig. 6 and Fig. 3.

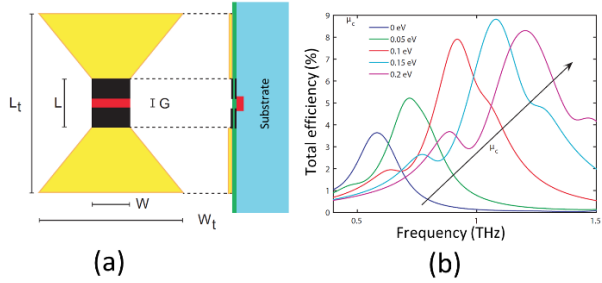

Fig. 7. Metal-graphene hybrid antenna [143]. (a) Top (left) and lateral (right) view of the structure. (b) Total radiation efficiency – considering a 10 kΩ photomixer as a source – of the hybrid antenna upon reconfiguration. © 2013 IEEE. Reprinted with permission, from [143].

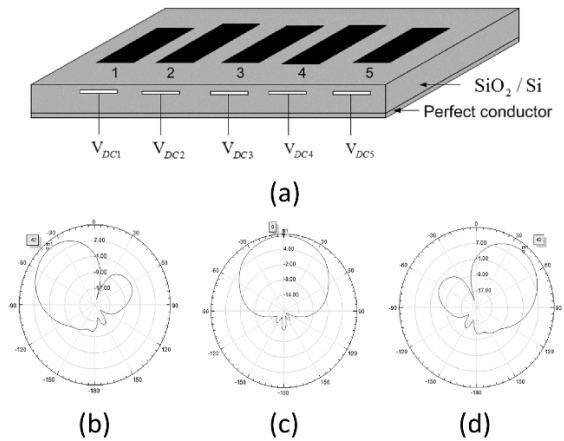

Fig. 8. Reconfigurable MIMO system based on graphene patch antennas [147]. (a) Schematic of the array. Each element is composed of a graphene layer with width $W=2$ $\mu m$ and length $L=5$ $\mu m$. (b)-(d) Radiation pattern of the antenna at 1 THz. (b) Elements 1, 3, and 4 are in a low-resistance state. (c) Elements 1, 3, and 5 are in a low-resistance state. (d) Elements 2, 3, and 5 are in low-resistance state. Graphene chemical potential and relaxation time are $\mu_c = 1.0$ eV and $\tau = 0.5$ ps, respectively. © 2014 IEEE. Reprinted with permission, from [147].

A different approach to exploit the tunability of graphene's conductivity in the design of THz antennas consists on using it as a perturbation in other structures with potentially higher radiation efficiency, typically involving metals [143]. The goal is to use graphene to electrically tune the resonant frequency, but not as the (main) radiating element. This concept is illustrated in Fig. 7 [137]. The graphene area (red color) located at the antenna center has the same geometry as the previously discussed planar dipoles, and is connected to the source. The yellow area represents metal, designed to maximize radiation efficiency while efficiently coupling to the graphene strips. We highlight that graphene's main role here is to serve as a matching network between the photomixer source – which presents a very high input impedance – and the metallic structure – which presents a high radiation efficiency. By tuning graphene's conductivity like in previous planar dipoles, the input impedance of the whole structure can be controlled in a wide range, tuning the total efficiency of the antenna – the radiation efficiency is barely affected since graphene is not the main radiator. Fig. 7b shows the antenna total efficiency, demonstrating the wide tunability and increased in efficiency compared to the graphene-only dipole of Fig. 3-4, obtained at the cost of significantly increased size. This example once again illustrates the recurring trade-off between miniaturization and efficiency in graphene antennas. We stress that many other works have also considered the use of graphene as a parasitic element, for instance using it as a tunable substrate [144], [145] or as a high-impedance surface [146].

All graphene resonant antennas discussed in this section can be used as part of more complex systems of arrays. As an example, Fig. 8 illustrates its use in a multiple input, multiple output (MIMO) configuration [147]. The system is comprised of several graphene nano-patch antennas whose resonant frequencies can be independently controlled. This configuration permits to easily achieve efficient and tunable MIMO responses at THz while keeping a very compact geometry and integrated, subwavelength elements. Fig. 8b-d illustrate the capabilities of the system to produce different radiation patters by changing the conductivity of individual patches in an array of just a few elements.

## IV. LEAKY-WAVE ANTENNAS

Leaky-wave antennas (LWAs) have been a very active topic of antenna research in the past decades due to their interesting beam-scanning properties [148]–[159]. They are based on the energy leaking towards free-space from an electromagnetic wave that is propagating through a waveguide and whose phase velocity is greater than the speed of light. Even though most LWAs have been designed to operate at micro and millimeter waves, their underlying physics is fundamental to explain exotic physical concepts such as extraordinary transmission [160], [161], electromagnetically induced transparency [162], or Cherenkov radiation [163]. Graphene is particularly suited for manipulating leaky waves at THz and infrared frequencies, allowing to implement reconfigurable structures and, more importantly, to achieve unprecedented magnet-less non-reciprocal responses.

One common and powerful type of leaky-wave antennas is based on the sinusoidal modulation of reactance surfaces. This concept was proposed by Prof. Oliner in the 50s [164] and has recently received significant attention at microwaves [165]–[168]. Essentially, the periodic modulation allows to express the fields along and on top the surface as an infinite sum of Floquet harmonics, some of which may lie within the light cone ($k < k_0$) and become leaky waves, causing the structure to operate as an antenna. The main parameters that control this type of antennas are the average reactance $X_s$, the amplitude of the modulation $M$, and the modulation period $p$. The reactance should be inductive to support confined TM waves that strongly interact with the modulated surface – its specific value determines the wave propagation constant and confinement. The modulation amplitude and period control the radiation rate and angle, respectively. The effective modulated surface reactance (assuming modulation and propagation along $y$) can thus be expressed as

$$X'_s = X_s[1 + M\sin(2\pi y/p)]. \qquad (4)$$

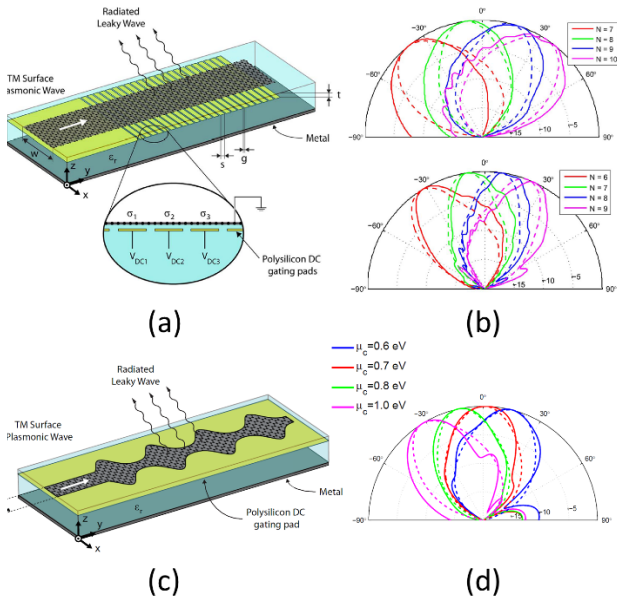

Fig. 9. Graphene leaky-wave antennas based on a sinusoidally modulated impedance [41], [42]. (a) 3D schematic of the first implementation, which is composed of a graphene sheet transferred onto a back-metallized substrate and several polysilicon DC gating pads located beneath it [42]. Spatial modulation is achieved through electrical gating. (b) Radiation diagram for different lengths of the electrically-induced structure periodic modulation [$\lambda_p = N \cdot (s + g)$]. Operation frequency is 2 THz, $g$=0.2 $\mu m$, and the applied voltage is in the $V_{DC} = 6.5 - 45$ V range. Top (bottom) diagrams have been generated considering an antenna transferred over a substrate with $\varepsilon_r = 3.8$ (1.8) and gating pads with $s$=4.8 (8.8) $\mu m$. (c) 3D schematic of the second implementation, which achieves the spatial modulation through the modulation of the graphene strip width [41]. (d) Radiation diagram at 1.5 THz for different chemical potentials. Results have been computed using full-wave simulations (solid lines) and leaky-wave theory (dashed lines). Graphene relaxation time is $\tau = 1.0$ ps. Panels (a), (b): © 2014 IEEE, reprinted with permission, from [42]. Panels (c), (d): © 2013 OSA. Reprinted with permission, [41].

Graphene is the perfect platform to implement this concept at THz frequencies, since it provides an inductive response, given by $Z_S = 1/\sigma = R_S + jX_S$, where $R_S$ and $X_S$ are the surface resistance and reactance of the sheet, respectively, that can be manipulated at will by exploiting graphene's field effect [98].

The pioneering first graphene-based LWA, proposed by Esquius-Morote and co-workers in 2014, is illustrated in Fig. 9a-b [42]. The structure is composed of a graphene sheet transferred onto a back-metallized substrate and a set of polysilicon DC bias pads located beneath it. By applying independent voltages to each pad, arbitrary modulations of the reactance profiles can be implemented over the sheet, thus allowing the propagation of leaky-waves. Importantly, the pointing angle $\theta_0$ and leakage rate $\alpha_{rad}$ of the radiated beam can be dynamically controlled by changing the features of the modulation. The voltage required in each pad depends on the substrate capacitance and the average reactance (average $\mu_c$), and can easily be found analytically [42]. The wavenumber $k_y$ of the fundamental harmonic along the modulated graphene sheet can be written as

$$k_y = \beta_{SPP} + \Delta\beta_{SPP} - j(\alpha_{SPP} + \alpha_{rad}) \quad (5)$$

where $\beta_{SPP}$ is the SPP propagation constant on the unmodulated sheet ($M$=0), $\Delta\beta_{SPP}$ is a small variation in the wavenumber due to the modulation (usually negligible), $\alpha_{SPP}$ the dissipative attenuation constant, and $\alpha_{rad}$ the attenuation constant due to leaky radiation. The space harmonic typically used for radiation in periodic leaky wave antennas is the -1 harmonic, with propagation constant $\beta_{-1} = k_z - 2\pi/p$. The pointing angle of the radiated beam is then

$$\theta_0 = \mathrm{asin}(\beta_{-1}/k_0) \approx \mathrm{asin}\left(\frac{\beta_{SPP}}{k_0} - \frac{\lambda_0}{p}\right). \quad (6)$$

Fig. 9b demonstrates outstanding capabilities of the antenna to switch from backward to forward radiation and quasi-broadside by simply changing the gate voltages. We do note that maximum radiation pointing exactly at broadside is not possible due to the open stopband of the periodic structure [148], [156]. The relatively lossy nature of graphene plasmons makes it challenging to simultaneously achieve highly directive beams and high-radiation efficiency, and one might decide to favor one over the other through the choice of the modulation amplitude: small $M$ for high effective aperture and thus higher directivity, or large $M$ for increased efficiency. We do stress, however, that this antenna, with realistic parameters of $\tau \approx 1$ ps and $M = 0.35$, has an impressive (for the THz band) radiation efficiency of roughly 11% while simultaneously providing beamscanning features.

A different implementation of graphene-based LWAs at THz was also proposed by same authors [41]. The antenna is shown in Fig. 9c-d and instead of gate pads uses a periodic modulation of the strip width to excite leaky spatial harmonics. To this purpose, authors take advantage of the large sensitivity of plasmons propagation constant with respect to the strip width. For ease of design and calculation, this structure can be modelled as an effective modulation of the reactance by simply mapping the width-dependence of the plasmon dispersion to effective changes on graphene's conductivity $\sigma = 1/Z_S$. Such analysis can efficiently be performed using the analytical scaling law of SPPs versus strips width introduced in [97], which was applied to realize quick designs and analysis of this type of antennas in [41]. This implementation is significantly simpler to fabricate than the previous one, since it avoids the use of gating pads beneath the graphene sheet, while still allowing a great control over the radiated beam direction at a fixed frequency by applying a single DC bias voltage (see Fig. 9d). There is, however, less control over the modulation amplitude $M$ and operation frequency than in the previous configuration, since they are determined by the width modulation at the fabrication stage.

Other approaches to implement LWAs include borrowing concepts from the microwave domain, such as composite right-left handed transmission lines and antennas [152], [153]. Specifically, [45] studied the viability of implementing such techniques at THz frequencies using graphene plasmonics. Even though the preliminary results are somewhat promising, the required structures are quite challenging to fabricate and need very high-quality graphene samples, which might delay its use in practice.

Elastic vibrations based on flexural (mechanical) waves have also been proposed as an alternative method to impart the periodic modulation required to excite leaky waves [44], fully exploiting the multidisciplinary nature of graphene. This approach, illustrated in Fig. 10, relies on the fact that a flexural wave travelling on graphene can be accurately modeled as a static grating with period $p$ (Δ in the figure). By changing the frequency of the flexural waves with a biharmonic source, the radiated beam can be dynamically controlled. This antenna has the main advantage of providing any desired modulation reactance profile, thanks to the flexibility of mechanical waves. In fact, its analog nature inherently avoids the discretization of the reactance profile using gate electrodes or its fixing during fabrication. Consequently, it can implement almost any desired radiation pattern and provide beamscanning functionalities. However, the structure is not fully planar and might be difficult to fabricate, since it requires free-standing graphene able to vibrate. Moreover, the adequate combination of miniaturized plasmonic devices and mechanical components may be challenging, especially in practical scenarios.

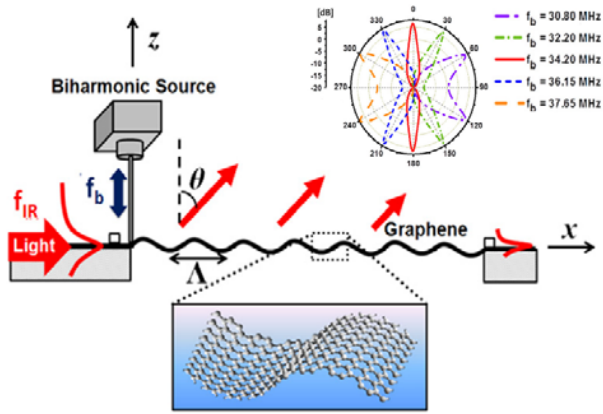

Fig. 10. Graphene leaky-wave antenna based on a sinusoidally modulated impedance achieved through elastic vibrations from flexural waves [44]. Inset (top right) shows the antenna radiation diagram at 30 THz versus different flexural frequencies. Graphene chemical potential and mobility are 0.5 eV and 20,000 cm²/V·s, respectively. © 2014 IOP Publishing. Reprinted with permission, from [44].

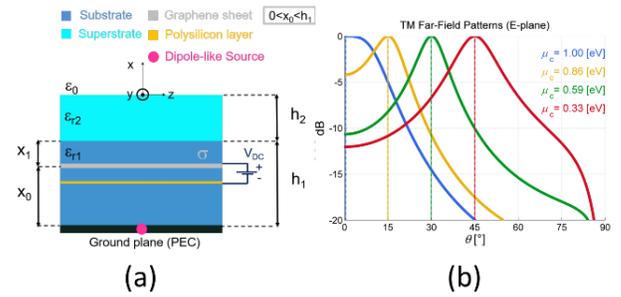

Fig. 11. Substrate-superstrate graphene-based leaky-wave antenna [170], [171]. (a) Cross section of the structure. A graphene sheet is located in the substrate, and a horizontal magnetic dipole is employed to excite the antenna from the ground plane. (b) Radiation patterns (E-plane) at 1.14 THz upon reconfiguration. Parameters: $\varepsilon_{r1} = 3.8$ (SiO₂), $\varepsilon_{r2} = 25$ (HfO₂), $x_0/h_1 = 0.81$, and graphene relaxation time is 0.5 ps. © 2016 IEEE. Reprinted with permission, from [171].

Another possibility to implement graphene-based LWAs is to use graphene as a perturbation element, as previously discussed for the case of resonant antennas. The motivation is, once again, to exploit the advantages of different technologies to maximize performance or reduce the restrictive dependence on high-quality graphene. In these designs, graphene is almost invariably used to provide quasi real-time electronic reconfigurability that would be almost impossible to achieve with conventional materials. This is the case of LWAs based on high-impedance surfaces [146], [169], which achieved enhanced beamforming and tunable features thanks to an adequate use of graphene. A recent example of an alternative approach is illustrated in Fig. 11, where graphene is included in a substrate-superstrate leaky wave antenna [170], [171]. Here, the position of the graphene sheet is optimized such that it significantly affects the propagation constant of the supported leaky modes. After a careful optimization, the structure exhibits strong reconfigurability, minimizes the leakage rate $\alpha_y$ to maximize directivity, and generates tunable radiation with nearly constant beamwidth. Fig. 11b demonstrates the reconfiguration capabilities of this antenna, with up to 45 deg of beam scanning by tuning the chemical potential from 1 eV to 0.3 eV using a single electrode.

In a different context, a recent and very promising trend in electromagnetics and in the antenna community is the pursuit of non-reciprocal responses without ferromagnetic materials or magneto-optic effects. This functionality traditionally requires bulky and expensive magnets that severely limit the miniaturization of devices and their integrability with other technologies. This is especially important in the case of THz graphene plasmonics, where the resulting components are significantly smaller than the silicon-incompatible required magnets. A new paradigm has recently been proposed to break time-reversal symmetry by modulating the properties of a structure (permittivity) in space and time. This approach has been successfully applied across the EM spectrum [56], [58], [59], [61], acoustics [55], [60], and even mechanics [172], allowing, for instance, the development of exciting circulators with state of the art performance that are compatible with integrated circuits [58]. A transition towards a magnet-less non-reciprocal paradigm will enable cheap and compact versions of essential devices such as isolators, circulators, and non-reciprocal antennas.

In the exciting work of Correas-Serrano [43] such concepts are translated into graphene plasmonics at THz by adequately exploiting the excellent reconfigurability of the material through its field effect. Specifically, this approach enables the development of THz antennas with different transmission and reception properties, as illustrated in Fig. 12. Even more interestingly, under time reversal, the SPPs transmitted/received by the device oscillate at different frequencies. The combination of these two phenomena give rise to unprecedented non-reciprocal responses in the context of LWAs. We note that relatively similar approaches have also been applied to develop non-reciprocal LWAs at microwaves [62], [173].

Such exciting responses can be achieved at THz simply using an antenna identical to the one of Fig. 9, but now making the bias signal in each pad oscillate at a frequency $f_m$, which creates different conductivity profiles along the direction $y$ at different instants, i.e. a spatiotemporal modulation of graphene conductivity, as shown in the inset of Fig. 12a [43]. This creates and *effective moving medium* for the propagating plasmons able to break time reversal symmetry (reciprocity), enabling for instance transmit-only or receive-only antennas. Keeping the same notation as in the static case, the surface reactance can now be expressed as

$$X'_S = X_S \left[ 1 + M \sin\left( \frac{2\pi}{p} y - 2\pi f_m t \right) \right]. \qquad (7)$$

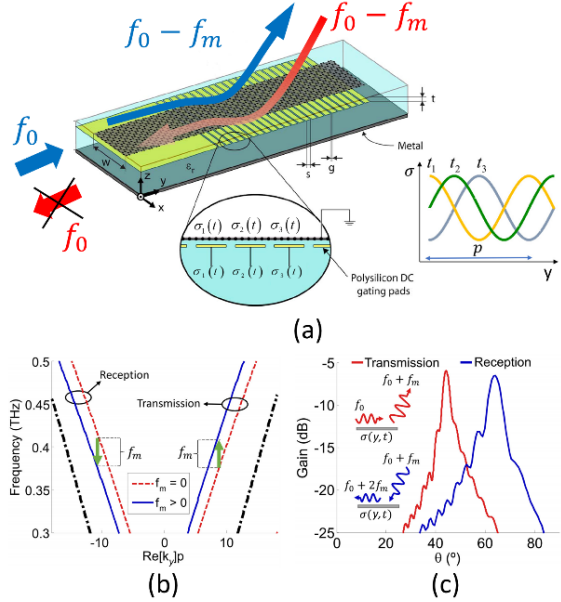

(a)

(b)

(c)

Fig. 12. Non-reciprocal graphene-based leaky-wave antenna based on a spatiotemporally modulated surface impedance [43], [46]. (a) 3D schematic of the structure. The antenna is like the one described in Fig. 9a, but now each gating pad is biased with a time-varying signal. The inset on the right illustrates the resulting graphene's conductivity profile, with a spatial period $p$ and a modulation frequency $f_m$, at instants $t_0$, $t_1=t_0+1/(4f_m)$ and $t_2=t_0+1/(2f_m)$. (b) Dispersion relation of the $n=-1$ harmonic supported by the antenna in transmission [propagation towards +y in panel a] and reception [towards -y]. Black dashed lines show the light cone. (c) Antenna gain in transmission and reception. Insets show the frequency conversion involved in both operation modes. Parameters are $f_0=350$ THz, $f_m=40$ GHz, graphene relaxation time $\tau = 0.5$ ps and unmodulated chemical potential $\mu_{c0} = 0.3$ eV. The structure is in free-space and the radiation angle $\theta$ is measured from broadside. © 2016 IEEE. Reprinted with permission, from [43], [46].

Importantly, this time-modulated surface reactance can easily be achieved at THz in moderately doped graphene ($|\mu_c| \gg k_B T$), a scenario where the *modulation of the chemical potential $\mu_c$ linearly translates to a similar modulation of graphene's conductivity* [43], [174]. In addition, graphene's conductivity can follow the gating signal variations up to hundreds of gigahertz [73], guaranteeing an adequate implementation of this approach. The dispersion relation of the modulated plasmons can be found analytically [43], [162], [164]. Fig. 12b shows the radiating harmonics for a spatiotemporally modulated case (blue, $f_m > 0$) and for static modulation (dashed red, $f_m = 0$). When $f_m > 0$, the radiating harmonic is upshifted (downshifted) in frequency by an amount equal to $f_m$ for propagation towards $+y$ ($-y$), implying different radiation properties in transmission and reception. Indeed, the result of the spatiotemporal modulation is an asymmetry in the dispersion diagram with respect to the frequency axis, which is a clear demonstration of the

nonreciprocal response. This exciting response can be better understood through Fig. 12c, which shows the gain of this LWA upon time-reversal, demonstrating up to 20 dB of contrast in transmission/reception for realistic graphene parameters, isolating the antenna. We stress again that the nonreciprocal response achieved is twofold: first, the gain and radiation pattern are significantly different in transmission and reception modes, and second, the SPPs transmitted/received oscillates at different frequencies under time reversal. Note that this non-reciprocal LWA maintains the beam scanning capabilities of its reciprocal counterpart, and it can electronically switch between standard reciprocal operation, transmit-only, and receive-only modes.

Given the recent interest by the engineering community in non-reciprocal devices and the ease of applying space-time modulations to graphene, we envision the development of exciting plasmonic antennas and guided devices based on this paradigm in the coming years.

## V. REFLECTARRAY ANTENNAS

Reflectarrays are a very well stablished topic in the RF, microwave, and antenna research community [175]–[177], and have enabled many terrestrial and satellite communication systems with outstanding capacity gain in the last decades [178]–[180]. These structures comprise an illuminating feed antenna and an array of reflective unit cells that each introduce a certain phase-shift upon reflection of a wave on the surface. The general concept is depicted in Fig. 13a [175]. The combined effect of all cells allows to engineer the phase fronts of the reflected beam, thus synthesizing highly directive far field beams in nearly any direction. It is worth noting that both reflectarrays and transmitarrays [181] (where the phase-shift is applied in transmission rather than in reflection [48]) are based on the same phenomenon as the so-called 'generalized Snell laws' that have inspired abundant research by the optics community in the past few years [181]–[186]. Reflectarrays combine the advantages of parabolic reflectors and phased arrays, forming narrow, reconfigurable beams through planar light-weight structures with simpler feeding networks than usual array antennas. Different technologies have been proposed to dynamically control the phase of the unit cell at micro and millimeter waves, such as semiconductor diodes [187]–[189], MEMS lumped elements [180], [190], [191], or liquid crystals [192], [193]. They are, however, not suitable for the THz band mainly due to size and loss. Graphene is a very promising candidate to fill this void thanks to its high inductance and strong electric field effect that provide miniaturization and the required tunability to synthesize any reflection phase.

The first graphene-based reflectarray, illustrated in Fig. 13b, was introduced by E. Carrasco et al [47], [48]. The proposed unit-cell is composed of a square graphene patch transferred onto a SiO2 substrate, as further described in Fig. 13b. In conventional reflectarray patches made of gold or noble metals, their resonance occurs when the size of the patch is comparable to a half wavelength of the wave in the effective media, which usually leads to large structures. This scenario is different in the case of graphene patches, since, as previously discussed in Section I, their resonant frequency is much lower than their metallic counterpart due to the plasmonic response of graphene. For instance, in the example of [48], the size of the required graphene patches is below λ/10, allowing to achieve an impressive miniaturization factor of around 5. Furthermore, a DC voltage applied between the graphene patch and the polycrystalline silicon layer allows to tune the patch reactance and thus the reflection phase of the unit cell. We stress that a thorough knowledge of the unit cell phase response, both versus its size or tunable features, is the most important step in any reflectarray design [175]. Fig. 13c shows the reflection phase (in degrees) for the tunable graphene patch versus chemical potential and patch size. Interestingly, the dynamic range of phase variation is much larger around low values of $\mu_c$ due to the dependence of the plasmon wavenumber on $\mu_c$, making them better suited for reflectarray design. This is qualitatively explained by Eq. 1, which shows that for large $\mu_c$ the plasmon wavenumber asymptotically approaches $k_0$. Once the reflection properties of the unit cell versus the design variables have been determined, the phase distribution that implements the desired far-field pattern can be readily found [47], [48]. In this case, the final reflectarray can be designed by varying both parameters (patch size and $\mu_c$) or keeping either of them constant. Fig. 13d shows the radiation pattern of a reflectarray based on size-modulation for $\mu_c = 0.0$ eV and $\mu_c = 0.29$ eV illustrating the excellent directivity and beam-scanning capabilities of this technology in the terahertz band.

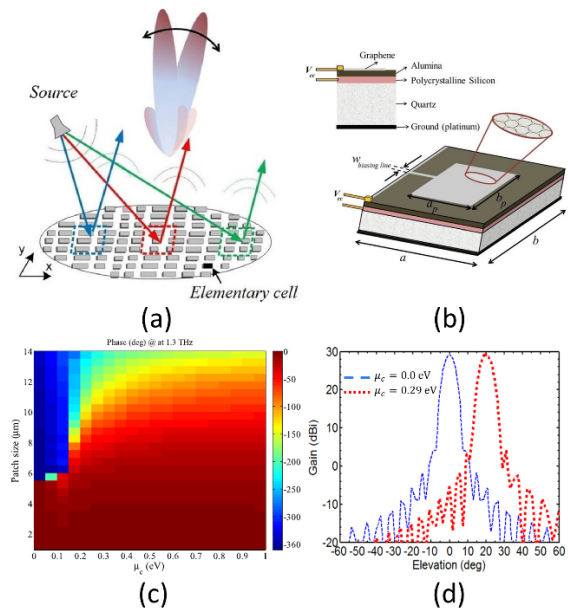

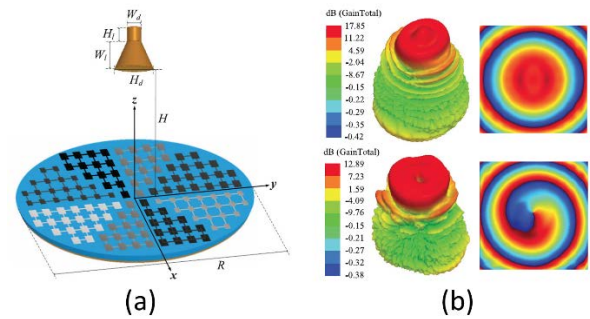

Fig. 14. Reconfigurable graphene-based reflectarray for vortex beam generation [206]. (a) 3D schematic of the structure. Dark and bright gray colors correspond to unit cells with different chemical potentials. (b) Simulated 3D radiation patterns (left) and phase front (right) of generated vortex beams with modes $l=0$ (top) and $l=-1$ (bottom). Details of the geometry and applied voltages can be found in [206]. © 2016 IEEE. Reprinted with permission, from [206].

Fig. 13. Reconfigurable graphene-based reflectarray [48]. (a) 3D schematic of the concept. (b) Details of a unit-cell element. (c) Phase shift (in degrees) induced in reflection at 1.3 THz by square patches of different sizes versus graphene chemical potential. (d) Antenna radiation diagram (E plane) at 1.3 THz upon reconfiguration. Graphene relaxation time is 1 ps and other parameters are $a=b=14$ μm, $a_p=b_p$ in the range of 0-10 μm, and quartz, alumina, and polysilicon thickness are 30, 0.1 and 0.05 μm, respectively. © Reprinted from [48], with the permission of AIP publishing.

One exciting application of reflectarrays is the generation of vortex beams [194]–[200]. Vortex beams are characterized by their orbital angular momentum (OAM), and they allow to increase the spectral efficiency thanks to the orthogonality of their ideally infinite number of states. Vortex beams have attracted significant attention in the past years thanks to promising experiments claiming that OAM might provide enhanced transmission capacity than any other known communication system [201]. Recently, it was clarified that OAM multiplexing is in fact a subset of MIMO systems [202], [203], and consequently, cannot bring larger capacity gain that them. Nevertheless, vortex beams can lead to exciting applications [204], for instance in communications with enhanced encryption capabilities, optical tweezers, optical trapping, LADAR (laser detection and ranging), and they propagate with less distortion than common Gaussian beams through turbulences. Unfortunately, the generation of vortex beams is a challenging task [196], [197], [205]. A common approach based on reflectarrays technology takes advantage of the reflection from a plane wave that impinges over a sectored circular surface, with each sector having the same reflection magnitude while their phase increments add up to a multiple $l$ of $2\pi$ ($l$ determines the order of the vortex beam). The required reflection coefficient of the $n$th sector and the corresponding surface impedance can be found analytically [206].

Since vortex beam generation requires precise control of the reflection phase, graphene patches offer once again a perfectly suited platform with large design flexibility. This concept is illustrated in Fig. 14 [199]. The structure is qualitatively like the previous reflectarray, but the patch parameters now divide the circular surface in phase sectors. Like before, the key step in the design is the characterization of reflection properties versus the unit cell parameters, which follows the same guidelines as discussed above and is omitted here for brevity [206]. In the pictorial representation of Fig. 14a, each shade of gray corresponds to patches with different $\mu_c$. Fig. 14b shows the radiation pattern and phase fronts of the generated vortex beams with modes $l = 0$ and $l = -1$, but we note that higher order modes can also be implemented by simply changing the biasing scheme. Importantly, this structure can also function as a regular beam-scanning reflectarray using appropriate bias values.

The same reflectarray concepts can be used to steer reflected or transmitted beams at will without necessarily implementing an antenna, akin to optics papers that exploit the generalized Snell's laws [182]. Fig. 15 implements light bending in reflection by means of an aperiodic array of graphene nanoribbons [49]. To design such device, it is essential to fully understand the reflection phase of individual strips, both as a function of its width and chemical potential, as investigated in Fig. 15b. Using this information, an appropriate selection of the strip widths permits to implement a linear phase discontinuity profile able to reflect an incident beam to the desired direction. Fig. 15c-d show the strip width distribution of the final reflectarray and a far-field radiation diagram, respectively. This configuration also provides tunable features, allowing to direct the reflected beam to almost any direction in the space by using a unique DC bias, thus greatly simplifying the complex biasing scheme of previous graphene-based reflectarray designs [47], [48]. Lastly, we stress that all these concepts can also be implemented in transmission through transmitarrays [181], [207], but they are usually less suited for antenna applications due to increased difficulty in fabricating the bias circuitry and the requirement of using multilayers to achieve wide phase control with minimal reflections [181].

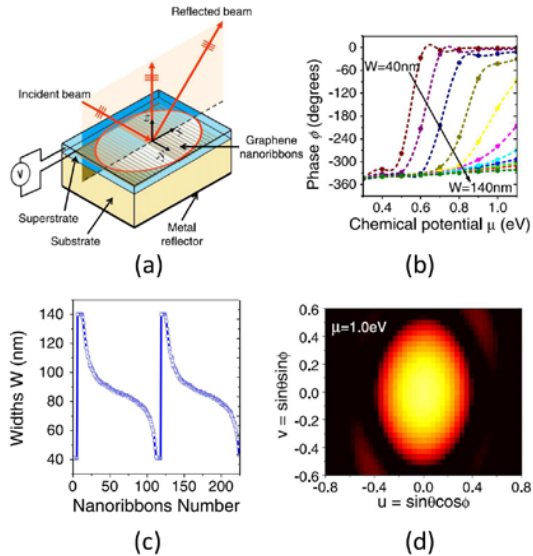

Fig. 15. Reconfigurable reflectarray based on aperiodic array of graphene ribbons [49]. (a) 3D schematic of the structure. (b) Phase of the reflection coefficient as a function of graphene chemical potential for different widths W of the nanoribbons. Mutual coupling between identical neighboring ribbons is considered. (c) Width of the nanoribbons across the device. (d) Far-field radiation pattern for $\mu = 1.0$ eV when a Gaussian beam illuminates the device at an incident angle of 45. The reflected beam can be steered by tuning $\mu$. Further details of the geometry can be found in [49]. © 2015 IOP Publishing. Reprinted with permission, [49].

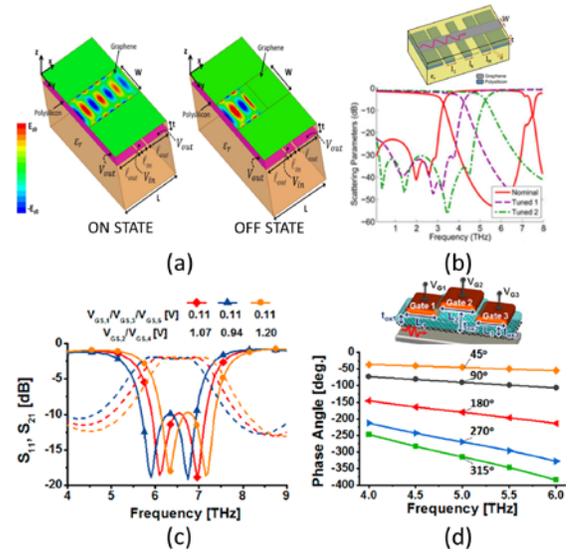

Fig. 16. Graphene-based components for THz antenna feeding networks. (a) Plasmonic switch based on graphene's field effect [213]. (b) Low-pass THz filter [97]. (c) Band-pass filter [214]. (d) Phase-shifters [215], [214]. Inset on the top of panels (a) and (d) describe potential implementations of the devices. Further details of the geometries can be found in the literature. Panel (a): © 2013 OSA. Reprinted with permission [213]. Panel (b): © 2014 IEEE. Reprinted with permission, from [97]. Panels (c), (d): 2014 American Chemical Society. Reprinted with permission, from [214].

## VI. FEEDING NETWORKS

Previous sections have exclusively discussed graphene-based antennas, i.e. structures able to couple guided energy to far field radiation (or vice versa). However, future THz transceiver front ends will also require guiding components such as interconnects, modulators, switches, filters, and phase shifters with high degrees of tunability enabled by graphene to properly feed the radiating elements in a unique, miniaturized, and multifunctional device. Clearly, an adequate interconnection between graphene devices and antennas will lead to lower losses and mismatch that using alternative technologies, plus it might also provide enhanced responses thanks to graphene's exotic properties. In this section, we review some of the most relevant advances in this area. The components described in the following exploit in some way the influence of graphene's chemical potential on the propagation constant and characteristic impedance of SPPs in graphene waveguides (see Eqs. 1 and 2), which allows to implement a very wide range of THz transmission lines by simply changing a biasing signal. This simple concept has tremendous implications, as virtually any textbook device that can be modeled using a transmission line approach can therefore be implemented using graphene plasmons in a miniaturized and reconfigurable THz platform [1], [25], [208]–[212].

Fig. 16 shows promising examples of several fundamental guided components, namely a switch [213], low-pass filter [94], band-pass filter [214] and a phase shifter [215], that are all based on the propagation of SPPs along graphene waveguides. The switch, illustrated in Fig. 16a, operates by introducing a strong impedance mismatch in a short graphene section with different chemical potential, easily obtained by biasing a gating pad located beneath that region. This simple approach allows to simultaneously achieve large isolation (>40 dB), low insertion loss, and fast switching speed thanks to graphene's field effect [73], [213], [216]. The low-pass and bandpass filters, shown in Fig. 16b-c, use a similar principle to implement traditional microwave filtering functions, where specific characteristic impedances and electrical lengths must be implemented at the desired frequency [214], [217], [218]. Specifically, each of the gated graphene sections can be modeled as a tunable transmission line, thus allowing to fully exploit decades of research in the 20th century dedicated to refining mathematical functions for optimal microwave filters [218]–[220]. Remarkably low insertion loss and band rejection are achieved in both structures, which also exhibit unprecedented electronic tuning of the filter central frequency. Fig. 16d illustrates a plasmonic phase shifter able to implement a wide range of phase shifts while providing small return loss, insertion loss, and phase error. These are just pioneer examples of potential guided devices, and that we expect many more microwave and optics devices to be successfully translated into THz graphene plasmonics soon.

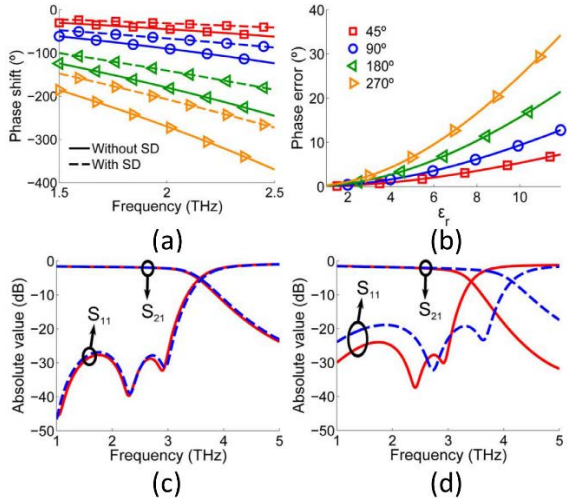

Fig. 17. Influence of spatial dispersion in the response of graphene-based phase shifters (top row) and low-pass filters (bottom row) [224]. The analysis of the phase shifter includes (a) phase difference between ports and (b) phase error due to spatial dispersion versus the permittivity of the surrounding media. The 7th-degree low pass filter is designed and analyzed in (c) free space and (d) embedded in Si ($\varepsilon_r \approx 11.8$). Solid lines: results neglecting spatial dispersion effects; dashed lines: results including spatial dispersion effects. Further details of the devices can be found in [224]. © 2014 IEEE. Reprinted with permission, from [224].

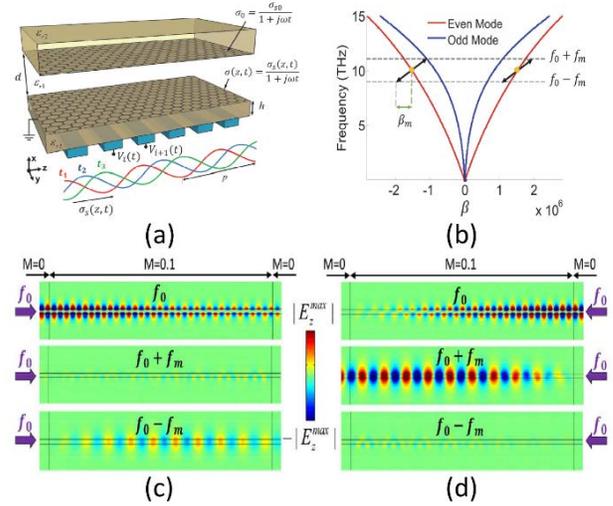

Fig. 18. Magnet-less non-reciprocal graphene guided device [43], [174]. (a) Schematic of a spatiotemporally modulated graphene waveguide. The conductivity of the bottom graphene layer is modulated through the time-varying voltages $V_i(t)$. The inset illustrates the resulting graphene's conductivity profile at different temporal instants. (b) Dispersion diagram of the unperturbed fundamental modes in a graphene-based parallel plate waveguide. Arrows indicate the frequency and wavenumber shifts imparted by the spatiotemporal modulation. (c)-(d) Electric field (longitudinal component) in the isolator at frequencies $f_0, f_0 \pm f_m$ when the device is excited at $f_0$ from (b) the left and (c) the right sides ($M$ is the modulation depth). Further details of the device can be found in [43]. © 2016 IEEE. Reprinted with permission, from [43], [174].

Usual analysis and design of graphene-based THz antennas and guided components have employed standard techniques borrowed from the microwave and optics communities [1], [2], [221]–[223]. However, there is a key difference when dealing with components based on ultra-confined surface plasmons polaritons (i.e., $k_{SPP} \gg k_0$, where $k_{SPP}$ and $k_0$ are the plasmon and free-space wavenumber, respectively) propagating in graphene: the influence of *nonlocal responses* or *spatial dispersion* [82], [96], [116], [209], [224]–[230]. This phenomenon is related to the finite Fermi velocity of electrons moving through the material and it fundamentally limits the maximum plasmon confinement. It arises because carriers on graphene have a fixed maximum speed ($v_F \approx c/300$), and therefore they cannot follow sharp spatial variation of ultra-confined surface plasmons, limiting the maximum SPPs wavenumber to be $k_{SPP} < 300k_0$ at any given frequency. In fact, spatial dispersion is non-negligible even for lower values, being significant in graphene when $k_{SPP} \gtrsim 100k_0$. We do stress that the spatial dispersion mentioned here – which effectively translates onto a wavenumber-dependent graphene conductivity $\sigma(k_{SPP})$ [109], [110] – is intrinsic to the band structure and properties of graphene, and should not be confused with the spatial dispersion that might arise in metamaterials due to the finite size of their composing unit-cells. Importantly, nonlocality might affect some devices more than others, depending on their features and operation principle. In the case of THz antennas, their radiating capabilities are always linked to their electric size in terms of free-space wavelength, which forces trade-offs between miniaturization and antenna figures of merit like gain. For this reason, ultra-confined SPPs are usually avoided in antennas, since their coupling to far-free radiation can be challenging. This is in principle not the case for guided devices. Their dimensions can eventually be reduced as much as desired, as long as in- and out-coupling of power can be done efficiently, so ultra-confined SPPs find a key role in these devices. Such SPPs can easily be obtained by decreasing the chemical potential, increasing the substrate permittivity (see Eq. (1)) of graphene waveguides, or using extremely anisotropic and hyperbolic configurations [231]–[235]. Thus, incorrectly assuming a local graphene response can lead to very significant errors in the design of guided component such as phase shifters and filters, as illustrated in Fig. 17 [224]. For instance, Fig. 17a-b show the phase error that may appear in graphene phase shifters due to spatial dispersion, demonstrating a strong dependence with the substrate used. Fig. 17c-d consider the influence of spatial dispersion in low-pass filters like those of Fig.16. In the free-standing case (Fig. 17c), the local conductivity model is a very good approximation, whereas performance is severely affected in Fig. 17d, which uses a high permittivity substrate. The operation frequency is shifted, return loss increased, roll-off decreased, and the implementation requires higher voltages. Similar issues are expected for extremely subwavelength plasmonic switches [213], [236]–[238] and other guided devices. In general, spatial dispersion limits the maximum tunable range of graphene plasmonic devices, reduces their quality factor, and degrades their frequency response [96], [116], [209], [224]. However, these effects can usually be considered and counteracted in the design of graphene-based devices by using accurate nonlocal models of graphene conductivity [116], [224].

Non-reciprocal waveguides are also of great importance for future THz devices to isolate sources from unwanted reflections or to enable full-duplex communications. Several groups have proposed non-reciprocal components based on magnetically biased graphene [52], [54], [111], [239]–[241], which may from significant drawbacks, namely the potential large size of the required magnets and their lack of compatibility with integrated circuits – in practical applications. A more promising solution for feeding networks is to adopt the magnet-less spatiotemporal modulation scheme described in Section III and applied there to develop non-reciprocal LWAs [43], [46]. This strategy might also allow to achieve non-reciprocal responses in both radiating elements and feeding networks, thus increasing isolation to almost any desired level. Fig. 18 illustrates the operation of a spatiotemporally modulated graphene parallel plate waveguide that implements a waveguide plasmonic isolator [43], [82]. By appropriately modulating the conductivity of one of the graphene layers in space and time, unidirectional mode conversion can be realized. Fig. 18b depicts the operation principle. It shows the dispersion of the two fundamental modes of the unmodulated structure, as well as the non-reciprocal wavenumber and frequency shifts induced by the modulation (black arrows). The modulation enables phase-matching and thus full exchange of energy in the backward direction (propagation towards -z), but not in the forward one (propagation towards +z). This isolating phenomenon is visually apparent in Fig. 18c-d. When the structure is excited from the left (towards +z), the modulation does not phase-match the modes, so only very small residual coupling occurs. When excited from the right (towards -z), phase-matching is perfect, and all power at the original frequency $f_0$ is converted to a different mode and frequency, isolating the ports. The residual power at $f_0 + f_m$ is entirely coupled to an orthogonal mode that can be readily scattered or reflected with modal filters. We stress that this pioneering example is just the first magnet-less non-reciprocal guided-device using THz graphene plasmonics, and we expect more refined configurations to be developed, such as integrated circulators, as well as theoretical studies on the fundamental limits of non-reciprocal responses.

## VII. APPLICATIONS

The most immediate application for graphene antennas is in THz transmitter front ends for communication systems. Successfully combining graphene antennas with passive plasmonic devices and efficient THz sources will lead to unprecedented wave manipulation functionalities in the THz band, enabling picocell and interchip wired or wireless communications with ultra-high bandwidth [16], [18], [19].

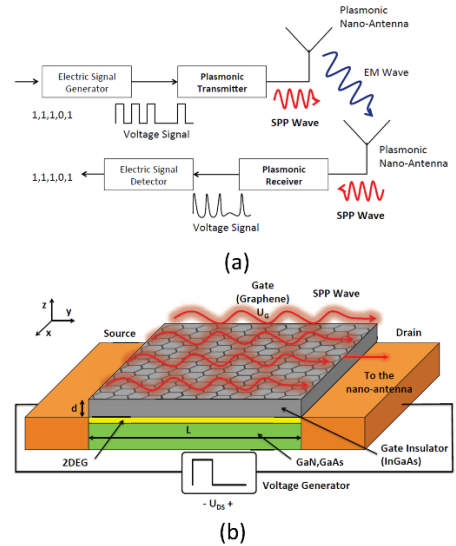

Fig. 19. Conceptual nano-transceiver architecture (a) and a specific implementation of a plasmonic nano-transmitter (b) based on high electron mobility transistor (HEMT) built with a III-V semiconductor and enhanced with graphene [242]. © 2014 IEEE. Reprinted with permission, from [242].

Although efficient excitation and detection of THz plasmons is still a challenging problem, significant strides have been made in parallel to the development of the passive devices that occupy this review. Excitation schemes used by most groups around the world do not currently employ graphene, using instead variations of conventional near-field techniques that have been shown to excite plasmons up to a hundred times shorter than the free-space wavelength [63], [64]. Nonetheless, some groups have proposed graphene-based sources that aim to maximize compatibility with graphene feeding networks and antennas with minimal interconnection [140], [242]. Fig. 19 shows an example of this approach based on a High Electron Mobility Transistor (HEMT) [242]. In transmission, the applied voltage between the drain and source accelerates a beam of electrons in the HEMT channel that excite a plasma wave able to couple to graphene SPPs. In reception, reciprocity allows to simply measure a drain-source voltage, more easily detected than near fields, especially in integrated systems. The graphene SPPs generated in this way can readily drive resonant and leaky-wave antennas, or any other passive device with good impedance matching.

Besides THz communications, one of the most promising applications of graphene plasmons is in biological sensing, since DNA molecules and many large proteins have rotational and vibrational modes in the THz and far-IR [17], [214], [243]–[249]. Plasmonic sensing has a long history at optical frequencies using noble metals like gold and silver [160], [250]–[254], but this technology could not be translated to THz before the advent of graphene. Moreover, the thinness and optical transparency of graphene make it very attractive for wearable and implantable sensors. The extreme field enhancement enabled by graphene plasmons in the THz band makes electromagnetic observables such as transmittance, phase, or polarization extremely sensitive to small perturbations in the complex permittivity of the surroundings [255], [256]. This sensitivity can be exploited to detect small concentrations of particles that respond strongly to THz light. Fig. 20a depicts a THz biosensor concept based in this principle, where the biological agents to be analyzed flow through a graphene plasmonic waveguide [214]. The presence of these molecules causes small changes in the channel's properties, detectable through the scattering parameters of the waveguide. This is especially suitable for molecules with THz absorption resonances, as they have a strong effect on the channel's permittivity. Note that, in this sensor, SPPs in the waveguide must be launched by either an external near-field source or an antenna. In contrast, Fig. 20b shows a recently proposed device where all the required elements (the sensor, frequency modulator, and antenna for energy harvesting) are included in a single module [257]. The structure is designed as a dual resonant antenna part of a graphene field effect transistor (GFET). The rectifying mechanism of the GFET generates a second harmonic signal whose intensity depends on the concentration of molecules to be sensed (this is a very sensitive process [257]). Since the monopole antenna is designed to resonate at both the original and the second harmonic frequency, the sensing output is automatically reradiated and available for processing.

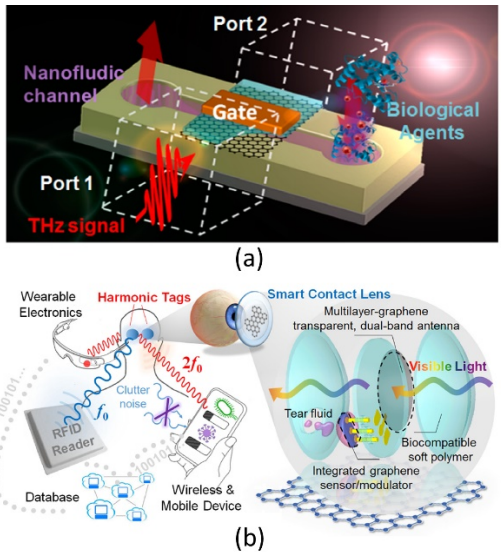

(a)

(b)

Fig. 20. Graphene-based THz bio-sensors. (a) Schematics of a THz biosensor composed of a nanofluidic delivery channel and the structure shown in Fig. 16c [214]. (b) Schematics of a graphene-based harmonic sensor designed for various point-of-care monitoring and wireless biomedical sensing [257]. The right panel illustrates an eye-wearable device (smart contact lens) based on the device, which may detect in real time the pathogen, bacteria, glucose, and infectious keratitis. Panel (a): © 2014 American Chemical Society. Reprinted with permission, from [214]. Panel (b) Reprinted from [257], with permission of AIP Publishing.

We envision future research on THz sensors to follow several directions: first, using metasurface concepts like hyperbolic dispersion to further enhance the near fields for increased sensitivity [117], [231], [232], [235], [258]–[264], second, designing and optimizing graphene nanoparticles supporting ultra-sensitive localized surface plasmon resonances (the most sensitive plasmonic nano-sensors at optics are based on nanoparticles [250]), third, using nonlinear responses [265]–[273], and fourth, employing nano-mechanical resonators [274], [275]. All of this must be accompanied by suitable instrumentation capable of high-throughput detection. In this context, graphene THz detectors have been subject of study for a few years, and promising advances have been made theoretically and experimentally. Current commercially available THz detectors are either based on thermal sensing elements or on nonlinear electronics. The former are typically slow or require cryogenic cooling, while the latter are limited to frequencies lower than 1 THz [276]–[278] . A more recent approach relies instead on field-effect transistors where a plasma wave is excited in the transistor channel by the impinging electric field, potentially enabling faster and more sensitive detection [279]–[283]. Different technologies have been used as platform for this class of FETs, including III-V high-electron-mobility transistors, Si-based MOS, and InAs nanowires [276], [278]. However, their performance is still inferior to other technologies, and is limited to the low THz. Graphene holds the promise to overcome these issues due to its higher electron mobility and suitability for manipulating THz waves. A recent example working at room temperature is depicted in Fig. 21 [70]. A log-periodic circular-toothed is patterned between the source and gate of a GFET. The concept is similar to the single-module sensor discussed above and the THz transceiver from Fig. 19, but the structure is now optimized for operation as a detector. When a THz oscillating electric field is fed between the gate and the channel of the GFET, the wave is rectified by the nonlinearity of the transfer function [70], inducing a DC signal between source and drain that is proportional to the received optical power. Fig. 21b plots the photoresponse versus incident THz power in fabricated GFETs, demonstrating voltages of the order of microvolts for incident powers of 1 miliwatt. The NEP of this detector is reported around ~30 nW Hz$^{-1/2}$, and the authors also demonstrate THz imaging with daily life objects [70] – THz waves are ideal for imaging common dielectric substances, hence their importance in security-related applications [17]. Terahertz imaging also has promise for label-free DNA analysis methods with increased diagnosis accuracy and lower cost, and in medical diagnosis for detection of anomalous tissue [17], [284]–[286]. There is still much room for improvement in GFET detectors, most notably by perfecting graphene fabrication processes, but the outlook is certainly promising, and this technology could soon have a real impact on society [71], [287]–[291]. For instance, sandwiching graphene between hexagonal boron nitride layers leads to significant improvements in mobility as has been recently demonstrated by several groups in other contexts [63], [68].

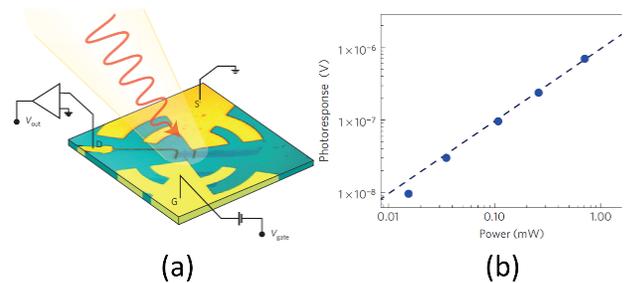

(a)                    (b)

Fig. 21. Graphene-based THz detector based on a log-periodic circular-toothed antenna [70]. (a) 3D schematic of the structure. Off-axis parabolic mirrors focus the THz beam into the device. The detector consists of a log-periodic circular-toothed antenna patterned between the source and gate of a GFET. The drain is a metal line running to the bonding pad (b) Photoresponse signal of the detectors a function of incident terahertz power. Dots are the experimental data; the dashed line is a fit to the data. Adapted by permission from Macmillan Publishers Ltd: Nature Materials [70], copyright 2012.

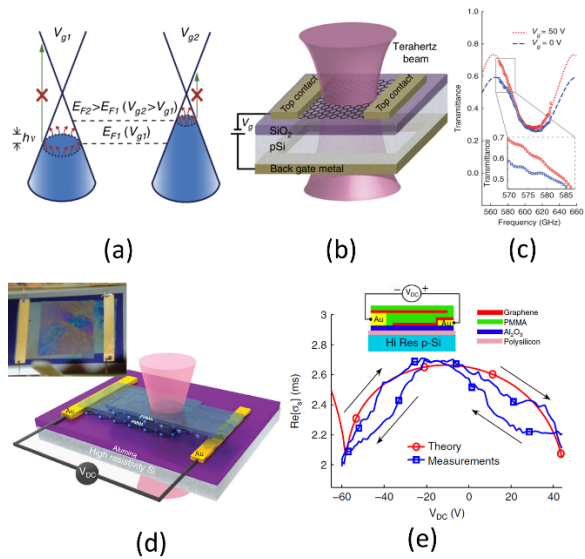

Fig. 22. Graphene-based THz modulators [72], [76]. (a) Conical band structure of graphene and associated optical processes. Intraband transitions (red arrows) dominate under terahertz illumination [72]. (b) Schematic of a single-layer graphene terahertz modulator. A monolayer graphene sheet was transferred on a $SiO_2$/p-Si substrate. (c) Measured (symbols) and modelled (lines) intensity transmittance for a single-layer modulator as a function of frequency for back gate voltages of 0 V (blue) and 50 V (red) [72]. (d) Schematic of a double-layer graphene modulator, where the layers gate each other [76]. Inset (top right) shows an illustration of the fabricated device. (e) Measured conductivity of the stack plotted versus a voltage $V_{DC}$ applied between the two graphene sheets [76]. Simulated results are included for comparison. Panels (a)-(c) adapted by permission from Macmillan Publishers Ltd: Nature Communications [72], copyright 2012. Panels (d),(e) adapted by permission from Macmillan Publishers Ltd: Nature Communications [76], copyright 2015.

The last major class of device we will consider, the modulator, will also be essential for a transition towards practical THz high speed communications. THz modulator technology is in its infancy, and until recently their development was lagging behind passive devices and sources/detectors [292]. Many attempts have been made at semiconductor modulators, but they always suffered from either low modulation frequency or low modulation depth, limited by the achievable tunability in carrier density. Graphene is the ideal material to overcome these two issues, since carriers can be injected very fast, holes conduct as well as electrons, and it interacts strongly with THz fields [26], [73]. As one may expect, the operation of graphene modulators is usually based on rapidly tuning the chemical potential using a dynamic electronic bias to modulate the conductivity and thus the transmission. Pioneering experiments by Sensale-Rodriguez et al demonstrated a few years ago substantial improvements over the state of the art in a simple modulator [72], [77], [293]. The structure, depicted in Fig. 22a-c, comprises a graphene monolayer transferred on a $SiO_2$/p-Si substrate and biased with metal gates. A voltage applied between the gates injects carriers on graphene, modulating the chemical potential. Fig. 22c plots the measured intensity transmittance as a function of frequency for back-gate voltages of 0 and 50 V, showing a 15% modulation depth at around 570 GHz. This performance was remarkable at the time, especially considering the simplicity of the design and the poor-quality graphene used in the experiment. A promising alternative to metal gates relies instead on self-biased graphene stacks where the tuning voltages are applied directly between graphene layers, as illustrated in Fig. 22d (inset shows the fabricated sample) [76], [127]. This approach minimizes the footprint and potentially reduces the RC constant of the structure, enabling faster and smaller devices. For small separation between the graphene sheets, the structure behaves as a single ultra-thin layer with conductivity equal sum of the two - this concept can be extended to an arbitrary number of layer pairs [72], [76]. The reconfiguration capabilities of the stack are plotted in Fig. 22e, demonstrating ambipolar tunability of the overall stack conductivity. Incorporating metal gates or additional graphene sheets to the modulator further boosts the available range of conductivity configurations, as discussed in [76]. Other modulators have been proposed and fabricated at THz using the same principles [74], [99], [292], and also at infrared frequencies using integrated waveguides [73], [75]. At infrared, graphene offers similar tunability and it can be easily integrated with the low-loss, technologically mature silicon photonics structures. In Ref. [73], the authors demonstrated modulation speeds of up to 30 GHz in hybrid dielectric-graphene waveguide modulators [73] using low quality graphene – improvements in graphene quality and low-resistance contacts may allow speeds in the hundreds of GHz [73]. Importantly, graphene's loss is less critical in this type of structure because its function is to perturb the photonic modes (with very high quality factor), rather than to be the main mechanism of energy propagation. These results are also a very strong indicator of the far-reaching, untapped potential of this hybrid approach, which we predict will be exploited in coming years not only to implement modulators, but also as an enabler of device tunability and non-reciprocal responses via spatiotemporal modulation in low-loss photonic technology.

## VIII. CONCLUSION AND OUTLOOK

We have reviewed the evolution of graphene antennas, from the inception of this field to the present state of the art, discussing their operation, shortcomings, and prospects. Two common themes stand out throughout this review: reconfigurability and miniaturization. These are the most unique and attractive features of graphene, and they are the cornerstone that has allowed researchers to conceive widely tunable antennas, ultra-fast modulators and detectors, sensitive plasmonic sensors, and magnet-less non-reciprocal plasmons. Although the practical realization of most of these devices has so far proved difficult, recent progress in graphene fabrication processes suggests a promising future. This is without a doubt the most important missing piece towards feasible graphene technology. From the conceptual point of view, we believe research will follow several directions in the coming years. First, we expect an increased cross-fertilization with the field of metamaterials and metasurfaces, exploiting the exotic electromagnetic phenomena that they enable. A second likely direction consists of using graphene's ultra-fast reconfigurability to introduce spatiotemporal modulations able to break time-reversal symmetry, enabling a wide variety of non-reciprocal devices. A third promising direction consists of combining graphene plasmonics with other technologies like silicon photonics to reduce the sometimes-stringent requirements on graphene quality. We envision, in the long term, that all these pieces will converge into a fully integrated, miniaturized, multifunctional, and reconfigurable graphene platform able not only to push forward but to transform current boundaries of THz technology.

REFERENCES

[1]     C. A. Balanis, *Antenna Theory: Analysis and Design*. John Wiley & Sons, 2005.

[2]     W. L. Stutzmann and G. A. Thiele, *Antenna Theory and Design*. John Wiley & Sons, 2012.

[3]     G. Maral and M. Bousquet, *Satellite communications systems: systems, techniques and technology*. John Wiley & Sons, 2011.

[4]     J. J. Spilker, *Digital communications by satellite*. Prentice-Hall, Inc., 1977.

[5]     R. J. Mailloux, *Phased array antenna handbook*, vol. 2. Artech House Boston, 2005.

[6]     R. C. Hansen, *Phased array antennas*, vol. 213. John Wiley & Sons, 2009.

[7]     R. Garg, *Microstrip antenna design handbook*. Artech house, 2001.

[8]     B. Widrow, P. E. Mantey, L. Griffiths, and B. Goode, "Adaptive antenna systems," *J. Acoust. Soc. Am.*, vol. 42, no. 5, pp. 1175–1176, 1967.

[9]     M. I. Skolnik, *Introduction to radar systems*. 1962.

[10]    J. C. Curlander and R. N. McDonough, *Synthetic aperture radar*. John Wiley & Sons New York, NY, USA, 1991.

[11]    R. G. Vaughan and J. B. Andersen, "Antenna diversity in mobile communications," *IEEE Trans. Veh. Technol.*, vol. 36, no. 4, pp. 149–172, 1987.

[12]    P. Soontornpipit, C. M. Furse, and Y. C. Chung, "Design of implantable microstrip antenna for communication with medical implants," *IEEE Trans. Microw. Theory Tech.*, vol. 52, no. 8, pp. 1944–1951, 2004.

[13]    W. Sun, G. J. Haubrich, and G. L. Dublin, "Implantable medical device microstrip telemetry antenna." Patent, 1999.

[14]    M. D. Amundson, J. A. Von Arx, W. J. Linder, P. Rawat, and W. R. Mass, "Circumferential antenna for an implantable medical device," 2002.

[15]    Y. Shimada, H. Iida, and M. Kinoshita, "Recent Research Trends of Terahertz Measurement Standards," *IEEE Trans. Terahertz Sci. Technol.*, vol. 5, no. 6, pp. 1166–1172, Nov. 2015.

[16]    P. H. Siegel, "Terahertz technology," *Microw. Theory Tech. IEEE Trans.*, vol. 50, no. 3, pp. 910–928, Mar. 2002.

[17]    B. Ferguson and X. Zhang, "Materials for terahertz science and technology.," *Nat. Mater.*, vol. 1, no. 1, pp. 26–33, Sep. 2002.

[18]    M. Tonouchi, "Cutting-edge terahertz technology," *Nat. Photonics*, vol. 1, no. 2, pp. 97–105, 2007.

[19]    P. H. Siegel, "Terahertz Technology in Biology and Medicine," *IEEE Trans. Microw. Theory Tech.*, vol. 52, no. 10, pp. 2438–2447, Oct. 2004.

[20]    P. H. Siegel, "THz instruments for space," *IEEE Trans. Antennas Propag.*, vol. 55, no. 11 I, pp. 2957–2965, Nov. 2007.

[21]    T. Nagatsuma, G. Ducournau, and C. C. Renaud, "Advances in terahertz communications accelerated by photonics," *Nat Phot.*, vol. 10, no. 6, pp. 371–379, 2016.

[22]    H. J. Yoon, J. H. Yang, Z. Zhou, S. S. Yang, M. M.-C. Cheng, and others, "Carbon dioxide gas sensor using a graphene sheet," *Sensors Actuators B Chem.*, vol. 157, no. 1, pp. 310–313, 2011.

[23]    K. Geim and K. S. Novoselov, "The rise of graphene," *Nat. Mater.*, vol. 6, no. 3, pp. 183–91, 2007.

[24]    A. N. Grigorenko, M Polini, and K S Novoselov, "Graphene plasmonics," *Nat. Photonics*, vol. 6, no. October, pp. 749–758, 2012.

[25]    M. Jablan, H. Buljan, and M. Soljačić, "Plasmonics in graphene at infrared frequencies," *Phys. Rev. B - Condens. Matter Mater. Phys.*, vol. 80, no. 24, p. 245435, Dec. 2009.

[26]    A. H. Castro Neto, F. Guinea, N. M. R. Peres, K. S. Novoselov, and A. K. Geim, "The electronic properties of graphene," *Rev. Mod. Phys.*, vol. 81, no. 1, pp. 109–162, 2009.

[27]    W. Huang, "Special issue on graphene," *Chinese Sci. Bull.*, vol. 57, no. 23, p. 2947, Nov. 2012.

[28]    C. Lee, X. Wei, J. W. Kysar, and J. Hone, "Measurement of the elastic Properties and Intrinsic Strength of Monolayer Graphene," *Science (80-. ).*, vol. 321, no. July, pp. 385–388, 2008.

[29]    D. C. Elias *et al.*, "Control of graphene's properties by reversible hydrogenation: evidence for graphane," *Science (80-. ).*, vol. 323, no. 5914, pp. 610–613, 2009.

[30]    C. Berger *et al.*, "Ultrathin Epitaxial Graphite: 2D Electron Gas Properties and a Route toward Graphene-based Nanoelectronics," *J. Phys. Chem. B*, vol. 108, no. 52, pp. 19912–19916, 2004.

[31]    A. A. Balandin, "Thermal properties of graphene and nanostructured carbon materials.," *Nat. Mater.*, vol. 10, no. 8, p. 569, 2011.

[32]    Y. Zhu *et al.*, "Graphene and graphene oxide: Synthesis, properties, and applications," *Adv. Mater.*, vol. 22, no. 35, pp. 3906–3924, 2010.

[33]    J. B. Pendry, L. Martin-Moreno, and F. J. Garcia-Vidal, "Mimicking Surface Plasmons with Structured Surfaces," *Science (80-. ).*, vol. 305, pp. 847–848, 2004.

[34]    J. M. Jornet and I. F. Akyildiz, "Graphene-Based Nano-Antennas for Electromagnetic Nanocommunications in the Terahertz Band," in *Proceedings of the Fourth European Conference on Antennas and Propagation (EuCAP)*, 2010, pp. 1–5.

[35] I. L. Martí, C. Kremers, A. Cabellos-Aparicio, J. M. Jornet, E. Alarcón, and D. N. Chigrin, "Scattering of terahertz radiation on a graphene-based nano-antenna," in *AIP Conference Proceedings*, 2011, vol. 1398, no. 1, pp. 144–146.

[36] J. S. Gomez-Diaz and J. Perruisseau-Carrier, "Microwave to THz properties of graphene and potential antenna applications," in *2012 International Symposium on Antennas and Propagation (ISAP)*, 2012, pp. 239–242.

[37] M. Tamagnone, J. S. Gómez-Díaz, J. R. Mosig, and J. Perruisseau-Carrier, "Analysis and design of terahertz antennas based on plasmonic resonant graphene sheets," *J. Appl. Phys.*, vol. 112, no. 11, 2012.

[38] M. Tamagnone, J. S. Gómez-Díaz, J. R. Mosig, and J. Perruisseau-Carrier, "Reconfigurable terahertz plasmonic antenna concept using a graphene stack," *Appl. Phys. Lett.*, vol. 101, no. 21, p. 214102, 2012.

[39] I. Llatser, C. Kremers, A. Cabellos-Aparicio, J. M. Jornet, E. Alarcón, and D. N. Chigrin, "Graphene-based nano-patch antenna for terahertz radiation," *Photonics Nanostructures - Fundam. Appl.*, vol. 10, no. 4, pp. 353–358, Oct. 2012.

[40] D. Correas-Serrano, J. S. Gomez-Diaz, A. Alu, and A. Alvarez-Melcon, "Electrically and Magnetically Biased Graphene-Based Cylindrical Waveguides: Analysis and Applications as Reconfigurable Antennas," *IEEE Trans. Terahertz Sci. Technol.*, vol. 5, no. 6, pp. 951–960, 2015.

[41] J. S. Gomez-Diaz, M. Esquius-Morote, and J. Perruisseau-Carrier, "Plane wave excitation-detection of non-resonant plasmons along finite-width graphene strips.," *Opt. Express*, vol. 21, no. 21, pp. 24856–72, Oct. 2013.

[42] M. Esquius-Morote, J. S. Gomez-Diaz, and Julien Perruisseau-Carrier, "Sinusoidally modulated graphene leaky-wave antenna for electronic beamscanning at THz," *IEEE Trans. Terahertz Sci. Technol.*, vol. 4, no. 1, pp. 116–122, 2014.

[43] D. Correas-Serrano, J. S. Gomez-Diaz, D. L. Sounas, Y. Hadad, A. Alvarez-Melcon, and A. Alu, "Nonreciprocal Graphene Devices and Antennas Based on Spatiotemporal Modulation," *IEEE Antennas Wirel. Propag. Lett.*, vol. 15, no. c, pp. 1529–1533, 2016.

[44] P.-Y. Chen, M. Farhat, A. N. Askarpour, M. Tymchenko, and A. Alu, "Infrared beam-steering using acoustically modulated surface plasmons over a graphene monolayer," *J. Opt.*, vol. 16, no. 9, p. 94008, Sep. 2014.

[45] D. A. Chu, P. W. C. Hon, T. Itoh, and B. S. Williams, "Feasibility of graphene CRLH metamaterial waveguides and leaky wave antennas," *J. Appl. Phys.*, vol. 120, no. 1, p. 13103, 2016.

[46] D. Correas-Serrano, A. Alvarez-Melcon, J. S. Gomez-Diaz, D. L. Sounas, and A. Alu, "Non-reciprocal leaky-wave antenna at THz based on spatiotemporally modulated graphene," in *2016 IEEE Antennas and Propagation Society International Symposium*, 2016, no. 1, pp. 1399–1400.

[47] E. Carrasco and J. Perruisseau-Carrier, "Reflectarray antenna at terahertz using graphene," *IEEE Antennas Wirel. Propag. Lett.*, vol. 12, pp. 253–256, 2013.

[48] E. Carrasco, M. Tamagnone, and J. Perruisseau-Carrier, "Tunable graphene reflective cells for THz reflectarrays and generalized law of reflection," *Appl. Phys. Lett.*, vol. 102, no. 10, pp. 1–5, 2013.

[49] E. Carrasco, M. Tamagnone, J. R. Mosig, T. Low, and J. Perruisseau-Carrier, "Gate-controlled mid-infrared light bending with aperiodic graphene nanoribbons array," *Nanotechnology*, vol. 26, no. 13, p. 134002, 2015.

[50] J. Perruisseau-Carrier, "Graphene for antenna applications: Opportunities and challenges from microwaves to THz," in *2012 Loughborough Antennas & Propagation Conference (LAPC)*, 2012, pp. 1–4.

[51] J. Perruisseau-Carrier, M. Tamagnone, J. S. Gomez-Diaz, and E. Carrasco, "Graphene antennas: Can integration and reconfigurability compensate for the loss?," *Eur. Microw. Week 2013*, pp. 369–372, 2013.

[52] D. L. Sounas and C. Caloz, "Electromagnetic nonreciprocity and gyrotropy of graphene," *Appl. Phys. Lett.*, vol. 98, no. 2, pp. 10–13, Jan. 2011.

[53] A. Fallahi *et al.*, "Manipulation of giant Faraday rotation in graphene metasurfaces," *Appl. Phys. Lett.*, vol. 101, no. 23, p. 231605, 2012.

[54] D. L. Sounas and C. Caloz, "Edge surface modes in magnetically biased chemically doped graphene strips," *Appl. Phys. Lett.*, vol. 99, no. 23, p. 231902, 2011.

[55] R. Fleury, D. L. Sounas, and A. Alu, "Subwavelength ultrasonic circulator based on spatiotemporal modulation," *Phys. Rev. B - Condens. Matter Mater. Phys.*, vol. 91, no. 17, p. 174306, 2015.

[56] N. A. Estep, D. L. Sounas, and A. Alù, "Magnetless microwave circulators based on spatiotemporally modulated rings of coupled resonators," *IEEE Trans. Microw. Theory Tech.*, vol. 64, no. 2, pp. 502–518, 2016.

[57] D. L. Sounas, C. Caloz, and A. Alu, "Giant non-reciprocity at the subwavelength scale using angular momentum-biased metamaterials.," *Nat. Commun.*, vol. 4, no. May, p. 2407, Jan. 2013.

[58] D. L. Sounas and A. Alù, "Angular-Momentum-Biased Nanorings to Realize Magnetic-Free Integrated Optical Isolation," *ACS Photonics*, vol. 1, no. 3, pp. 198–204, 2014.

[59] S. Qin, S. Member, Q. Xu, Y. E. Wang, and S. Member, "Nonreciprocal Components With Distributedly Modulated Capacitors," vol. 62, no. 10, pp. 2260–2272, 2014.

[60] W. F. Bottke, R. J. Walker, J. M. D. Day, D. Nesvorny, and L. Elkins-tanton, "Sound Isolation and Giant Linear Nonreciprocity in a Compact Acoustic Circulator," *Science (80-. ).*, vol. 343, no. 6170, pp. 516–519, 2014.

[61] N. A. Estep, D. L. Sounas, J. Soric, and A. Alu, "Magnetic-free non-reciprocity and isolation based on parametrically modulated coupled-resonator loops,"

*Nat Phys*, vol. 10, no. 12, pp. 923–927, Nov. 2014.

[62] Y. Hadad, J. C. Soric, and A. Alu, "Breaking temporal symmetries for emission and absorption," *Proc. Natl. Acad. Sci.*, vol. 113, no. 13, pp. 3471–3475, 2016.

[63] A. Woessner *et al.*, "Highly confined low-loss plasmons in graphene–boron nitride heterostructures," *Nat. Mater.*, vol. 14, no. 4, pp. 421–425, 2014.

[64] J. Chen *et al.*, "Optical nano-imaging of gate-tunable graphene plasmons.," *Nature*, vol. 487, no. 7405, pp. 77–81, 2012.

[65] K. J. Tielrooij *et al.*, "Generation of photovoltage in graphene on a femtosecond timescale through efficient carrier heating," *Nat. Nanotechnol.*, vol. 10, no. 5, pp. 437–443, 2015.

[66] V. W. Brar *et al.*, "Hybrid surface-phonon-plasmon polariton modes in graphene/monolayer h-BN heterostructures," *Nano Lett.*, vol. 14, no. 7, pp. 3876–3880, 2014.

[67] C. R. Dean *et al.*, "Boron nitride substrates for high-quality graphene electronics.," *Nat. Nanotechnol.*, vol. 5, no. 10, pp. 722–726, Oct. 2010.

[68] W. Gannett, W. Regan, K. Watanabe, T. Taniguchi, M. F. Crommie, and A. Zettl, "Boron nitride substrates for high mobility chemical vapor deposited graphene," *Appl. Phys. Lett.*, vol. 98, no. 24, pp. 99–102, 2011.

[69] P. J. Zomer, S. P. Dash, N. Tombros, and B. J. Van Wees, "A transfer technique for high mobility graphene devices on commercially available hexagonal boron nitride," *Appl. Phys. Lett.*, vol. 99, no. 23, pp. 99–102, 2011.

[70] L. Vicarelli *et al.*, "Graphene field-effect transistors as room-temperature terahertz detectors," *Nat. Mater.*, vol. 11, no. 9, pp. 1–7, Oct. 2012.

[71] F. Xia, T. Mueller, Y. Lin, A. Valdes-Garcia, and P. Avouris, "Ultrafast graphene photodetector," *Nat. Nanotechnol.*, vol. 4, no. 12, pp. 839–843, 2009.

[72] B. Sensale-Rodriguez *et al.*, "Broadband graphene terahertz modulators enabled by intraband transitions.," *Nat. Commun.*, vol. 3, p. 780, 2012.

[73] C. T. Phare, Y. D. Lee, J. Cardenas, and M. Lipson, "Graphene electro-optic modulator with 30 GHz bandwidth," *Nat. Photonics*, vol. 9, no. 8, pp. 511–514, 2015.

[74] R. Zhang *et al.*, "Broadband black phosphorus optical modulator in visible to mid-infrared spectral range," p. arXiv 1505.05992, 2015.

[75] M. Liu *et al.*, "A graphene-based broadband optical modulator.," *Nature*, vol. 474, no. 7349, pp. 64–67, 2011.

[76] J. S. Gomez-Diaz *et al.*, "Self-biased reconfigurable graphene stacks for terahertz plasmonics," *Nat. Commun.*, vol. 6, no. 6334, pp. 1–8, May 2015.

[77] B. Sensale-Rodriguez *et al.*, "Unique prospects for graphene-based terahertz modulators," *Appl. Phys. Lett.*, vol. 99, no. 11, pp. 1–4, 2011.

[78] B. Sensale-Rodriguez *et al.*, "Terahertz imaging employing graphene modulator arrays.," *Opt. Express*, vol. 21, no. 2, pp. 2324–30, 2013.

[79] A. S. Rodin, A. Carvalho, and A. H. Castro Neto, "Strain-induced gap modification in black phosphorus," *Phys. Rev. Lett.*, vol. 112, no. 17, pp. 1–5, May 2014.

[80] T. Low *et al.*, "Tunable optical properties of multilayers black phosphorus," *Phys. Rev. B - Condens. Matter Mater. Phys.*, vol. 75434, no. 7, pp. 1–5, Aug. 2014.

[81] J. Lu, J. Yang, A. Carvalho, H. Liu, Y. Lu, and C. H. Sow, "Light-Matter Interactions in Phosphorene," *Acc. Chem. Res.*, vol. 49, no. 9, pp. 1806–1815, 2016.

[82] D. Correas-Serrano, J. S. Gomez-Diaz, A. A. Melcon, and A. Alù, "Black phosphorus plasmonics: anisotropic elliptical propagation and nonlocality-induced canalization," *J. Opt.*, vol. 18, no. 10, p. 104006, 2016.

[83] D. Correas-Serrano, A. Alvarez-Melcon, J. S. Gomez-Diaz, and A. Alu, "Strong Light Matter Interactions in Thin Black Phosphorus Films," in *IEEE Antenna and Propagation Symposium, Fajardo (Puerto Rico)*, 2016.

[84] N. Morell *et al.*, "High Quality Factor Mechanical Resonators Based on WSe2 Monolayers," *Nano Lett.*, vol. 16, no. 8, pp. 5102–5108, 2016.

[85] K. F. Mak, C. Lee, J. Hone, J. Shan, and T. F. Heinz, "Atomically thin MoS2: A new direct-gap semiconductor," *Phys. Rev. Lett.*, vol. 105, no. 13, p. 136805, Sep. 2010.

[86] A. Splendiani *et al.*, "Emerging photoluminescence in monolayer MoS2," *Nano Lett.*, vol. 10, no. 4, pp. 1271–1275, Apr. 2010.

[87] E. Cappelluti, R. Roldan, J. A. Silva-Guillen, P. Ordejon, and F. Guinea, "Tight-binding model and direct-gap/indirect-gap transition in single-layer and multilayer MoS2," *Phys. Rev. B - Condens. Matter Mater. Phys.*, vol. 88, no. 7, 2013.

[88] B. H. Nguyen and V. H. Nguyen, "Two-dimensional hexagonal semiconductors beyond graphene," *Adv. Nat. Sci. Nanosci. Nanotechnol.*, vol. 7, no. 4, p. 43001, 2016.

[89] W. Zhao *et al.*, "Evolution of electronic structure in Atomically Thin Sheets of WS2 and WSe2," *J. Phys. Chem. Solids*, vol. 72, no. 5, pp. 474–478, Jan. 2011.

[90] Z. Fei *et al.*, "Electronic and plasmonic phenomena at graphene grain boundaries.," *Nat. Nanotechnol.*, vol. 8, no. 11, pp. 821–5, 2013.

[91] E. H. Hwang, S. Adam, and S. Das Sarma, "Carrier transport in two-dimensional graphene layers," *Phys. Rev. Lett.*, vol. 98, no. 18, pp. 2–5, 2007.

[92] L. A. Falkovsky and A. A. Varlamov, "Space-time dispersion of graphene conductivity," *Eur. Phys. J. B*, vol. 56, no. 4, pp. 281–284, 2007.

[93] K. S. Novoselov, "A roadmap for graphene," *Nature*, vol. 490, no. 7419, pp. 192–200, 2013.

[94] S. Bagheri, N. Mansouri, and E. Aghaie, "Phosphorene: A new competitor for graphene," *Int. J. Hydrogen Energy*, vol. 41, no. 7, pp. 4085–4095, 2016.

[95] G. W. Hanson, "Dyadic Green's functions and guided surface waves for a surface conductivity model of

graphene," *J. Appl. Phys.*, vol. 103, no. 6, p. 64302, 2008.

[96] D. Correas-Serrano, J. S. Gomez-Diaz, J. Perruisseau-Carrier, and A. Alvarez-Melcon, "Spatially dispersive graphene single and parallel plate waveguides: Analysis and circuit model," *IEEE Trans. Microw. Theory Tech.*, vol. 61, no. 12, pp. 4333–4344, Dec. 2013.

[97] D. Correas-Serrano, J. S. Gomez-Diaz, J. Perruisseau-Carrier, and A. Alvarez-Melcon, "Graphene-based plasmonic tunable low-pass filters in the terahertz band," *IEEE Trans. Nanotechnol.*, vol. 13, no. 6, pp. 1145–1153, Nov. 2014.

[98] K. S. Novoselov *et al.*, "Electric Field Effect in Atomically Thin Carbon Films," *Science,* vol. 306, no. 5696, pp. 666–669, 2004.

[99] W. Li *et al.*, "Ultrafast all-optical graphene modulator," *Nano Lett.*, vol. 14, no. 2, pp. 955–959, 2014.

[100] V. Lukose, R. Shankar, and G. Baskaran, "Novel electric field effects on landau levels in graphene," *Phys. Rev. Lett.*, vol. 98, no. 11, p. 116802, 2007.

[101] Y. Zhang *et al.*, "Landau-level splitting in graphene in high magnetic fields," *Phys. Rev. Lett.*, vol. 96, no. 13, p. 136806, 2006.

[102] M. L. Sadowski, G. Martinez, M. Potemski, C. Berger, and W. A. De Heer, "Landau level spectroscopy of ultrathin graphite layers," *Phys. Rev. Lett.*, vol. 97, no. 26, p. 266405, 2006.

[103] M. O. Goerbig, "Electronic properties of graphene in a strong magnetic field," *Rev. Mod. Phys.*, vol. 83, no. 4, p. 1193, 2011.

[104] K. Nomura and A. H. MacDonald, "Quantum Hall ferromagnetism in graphene," *Phys. Rev. Lett.*, vol. 96, no. 25, p. 256602, 2006.

[105] F. Guinea, A. H. C. Neto, and N. M. R. Peres, "Electronic states and Landau levels in graphene stacks," *Phys. Rev. B*, vol. 73, no. 24, p. 245426, 2006.

[106] Y. Zhang, Y.-W. Tan, H. L. Stormer, and P. Kim, "Experimental observation of the quantum Hall effect and Berry's phase in graphene," *Nature*, vol. 438, no. 7065, pp. 201–204, 2005.

[107] K. S. Novoselov *et al.*, "Unconventional quantum Hall effect and Berry's phase of 2pi in bilayer graphene," *Nat. Phys.*, vol. 2, no. 3, pp. 177–180, 2006.

[108] K. S. Novoselov *et al.*, "Room-Temperature Quantum Hall Effect in Graphene," *Science (80-. ).*, vol. 315, no. 5817, pp. 1379–1379, 2007.

[109] C. L. Kane and E. J. Mele, "Quantum spin Hall effect in graphene," *Phys. Rev. Lett.*, vol. 95, no. 22, p. 226801, 2005.

[110] D. L. Sounas and C. Caloz, "Graphene-based non-reciprocal spatial isolator," *IEEE Antennas Propag. Soc. AP-S Int. Symp.*, pp. 1597–1600, 2011.

[111] D. L. Sounas and C. Caloz, "Gyrotropy and nonreciprocity of graphene for microwave applications," *IEEE Trans. Microw. Theory Tech.*, vol. 60, no. 4, pp. 901–914, Apr. 2012.

[112] Z. Sun *et al.*, "Graphene Mode-Locked Ultrafast

Laser," *ACS Nano*, vol. 4, no. 2, pp. 803–810, 2010.

[113] S. Boubanga-Tombet, S. Chan, T. Watanabe, A. Satou, V. Ryzhii, and T. Otsuji, "Ultrafast carrier dynamics and terahertz emission in optically pumped graphene at room temperature," *Phys. Rev. B - Condens. Matter Mater. Phys.*, vol. 85, no. 3, pp. 3–8, 2012.

[114] V. Ryzhii, A. A. Dubinov, V. Y. Aleshkin, M. Ryzhii, and T. Otsuji, "Injection terahertz laser using the resonant inter-layer radiative transitions in double-graphene-layer structure," *Appl. Phys. Lett.*, vol. 103, no. 16, pp. 10–14, 2013.

[115] D. Svintsov, Z. Devizorova, T. Otsuji, and V. Ryzhii, "Plasmons in tunnel-coupled graphene layers: Backward waves with quantum cascade gain," *Phys. Rev. B - Condens. Matter Mater. Phys.*, vol. 94, no. 11, 2016.

[116] G. Lovat, G. W. Hanson, R. Araneo, and P. Burghignoli, "Semiclassical spatially dispersive intraband conductivity tensor and quantum capacitance of graphene," *Phys. Rev. B - Condens. Matter Mater. Phys.*, vol. 87, no. 11, p. 115429, Mar. 2013.

[117] D. Correas-Serrano, J. S. Gomez-Diaz, M. Tymchenko, and A. Alù, "Nonlocal response of hyperbolic metasurfaces," *Opt. Express*, vol. 23, no. 23, p. 29434, 2015.

[118] F. J. García de Abajo, "Graphene Plasmonics: Challenges and Opportunities," *ACS Photonics*, vol. 1, no. 3, pp. 135–152, 2014.

[119] S. Abadal, J. M. Jornet, I. Llatser, A. Cabellos-aparicio, E. Alarcón, and I. F. Akyildiz, "Wireless Nanosensor Networks using Graphene-based Nano-Antennas," p. 2011, 2011.

[120] O. V. Shapoval, J. S. Gomez-Diaz, J. Perruisseau-Carrier, J. R. Mosig, and A. I. Nosich, "Integral equation analysis of plane wave scattering by coplanar graphene-strip gratings in the thz range," *IEEE Trans. Terahertz Sci. Technol.*, vol. 3, no. 5, pp. 666–674, 2013.

[121] Y. R. Padooru, A. B. Yakovlev, C. S. R. Kaipa, G. W. Hanson, F. Medina, and F. Mesa, "Dual capacitive-inductive nature of periodic graphene patches: Transmission characteristics at low-terahertz frequencies," *Phys. Rev. B*, vol. 87, no. 11, p. 115401, Mar. 2013.

[122] C. S. R. Kaipa *et al.*, "Enhanced transmission with a graphene-dielectric microstructure at low-terahertz frequencies," *Phys. Rev. B*, vol. 85, no. 24, p. 245407, Jun. 2012.

[123] P. H. Siegel, P. De Maagt, and a. I. I. Zaghloul, "Antennas for terahertz applications," *2006 IEEE Antennas Propag. Soc. Int. Symp.*, pp. 2383–2386, 2006.

[124] M. Tamagnone and J. Perruisseau-Carrier, "Predicting Input Impedance and Efficiency of Graphene Reconfigurable Dipoles Using a Simple Circuit Model," *IEEE Antennas Wirel. Propag. Lett.*, vol. 13, no. c, pp. 313–316, 2014.

[125] D. Correas-Serrano, J. S. Gomez-Diaz, and A.

Alvarez-Melcon, "Surface plasmons in graphene cylindrical waveguides," *IEEE Antennas Propag. Soc. AP-S Int. Symp.*, vol. 0, no. 1, pp. 896–897, 2014.

[126] B. Zhu, G. Ren, Y. Gao, Y. Yang, Y. Lian, and S. Jian, "Graphene-coated tapered nanowire infrared probe: a comparison with metal-coated probes.," *Opt. Express*, vol. 22, no. 20, pp. 24096–103, Sep. 2014.

[127] D. Correas-Serrano, A. Alvarez-Melcon, J. S. Gomez-Diaz, and A. Alu, "Surface plasmon modes in self-biased coupled graphene-coated wires," *IEEE Antennas Propag. Soc. AP-S Int. Symp.*, vol. 2015–Octob, pp. 1638–1639, 2015.

[128] I. Soto Lamata, P. Alonso-González, R. Hillenbrand, and A. Y. Nikitin, "Plasmons in cylindrical 2D materials as a platform for nanophotonic circuits," *ACS Photonics*, vol. 2, no. 2, pp. 280–286, 2015.

[129] G. W. Hanson, "Fundamental transmitting properties of carbon nanotube antennas," *IEEE Trans. Antennas Propag.*, vol. 53, no. 11, pp. 3426–3435, 2005.

[130] P. J. Burke, S. Li, and Z. Yu, "Quantitative theory of nanowire and nanotube antenna performance," *IEEE Trans. Nanotechnol.*, vol. 5, no. 4, pp. 314–334, Jul. 2006.

[131] Z. Fang, Z. Liu, Y. Wang, P. M. Ajayan, P. Nordlander, and N. J. Halas, "Graphene-antenna sandwich photodetector," *Nano Lett.*, vol. 12, no. 7, pp. 3808–3813, 2012.

[132] X. Qin, J. Chen, C. Xie, N. Xu, and J. Shi, "A Tunable THz Dipole Antenna Based On Graphene," no. 4, pp. 5–7, 2017.

[133] M. Dragoman *et al.*, "A tunable microwave slot antenna based on graphene," *Appl. Phys. Lett.*, vol. 106, no. 15, p. 153101, 2015.

[134] K. Y. Shin, J. Y. Hong, and J. Jang, "Micropatterning of graphene sheets by inkjet printing and its wideband dipole-antenna application," *Adv. Mater.*, vol. 23, no. 18, pp. 2113–2118, 2011.

[135] D. Kim, "A High Performance IBC-Hub Transceiver for IntraBody Communication System," *Microw. Opt. Technol. Lett.*, vol. 54, no. 12, pp. 2781–2784, 2012.

[136] Z. Zhu, S. Joshi, S. Grover, and G. Moddel, "Graphene geometric diodes for terahertz rectennas," *J. Phys. D. Appl. Phys.*, vol. 46, no. 18, p. 185101, 2013.

[137] J. M. Jornet and A. Cabellos, "On the feeding mechanisms for graphene-based THz plasmonic nano-antennas," *IEEE-NANO 2015 - 15th Int. Conf. Nanotechnol.*, pp. 168–171, 2016.

[138] A. Cabellos-Aparicio *et al.*, "Use of Terahertz Photoconductive Sources to Characterize Tunable Graphene RF Plasmonic Antennas," *IEEE Trans. Nanotechnol.*, vol. 14, no. 2, pp. 390–396, 2015.

[139] R. Filter, M. Farhat, M. Steglich, R. Alaee, C. Rockstuhl, and F. Lederer, "Tunable graphene antennas for selective enhancement of THz-emission," *Opt. Express*, vol. 21, no. 3, pp. 3737–3745, 2013.

[140] P.-Y. Chen and A. Alù, "A terahertz photomixer based on plasmonic nanoantennas coupled to a graphene emitter.," *Nanotechnology*, vol. 24, no. 45, p. 455202, Oct. 2013.

[141] M. Tamagnone and J. R. Mosig, "Theoretical Limits on the Efficiency of Reconfigurable and Nonreciprocal Graphene Antennas," *IEEE Antennas Wirel. Propag. Lett.*, vol. 15, pp. 1549–1552, 2016.

[142] M. Tamagnone, A. Fallahi, J. R. Mosig, and J. Perruisseau-Carrier, "Fundamental limits and near-optimal design of graphene modulators and non-reciprocal devices," *Nat. Photonics*, vol. 8, no. 7, pp. 556–563, 2014.

[143] M. Tamagnone, J. S. G. Diaz, J. Mosig, and J. Perruisseau-Carrier, "Hybrid graphene-metal reconfigurable terahertz antenna," *IEEE MTT-S Int. Microw. Symp. Dig.*, pp. 9–11, 2013.

[144] M. Dragoman, A. A. Muller, D. Dragoman, F. Coccetti, and R. Plana, "Terahertz antenna based on graphene," *J. Appl. Phys.*, vol. 107, no. 10, pp. 10–13, 2010.

[145] Y. Yao *et al.*, "Broad electrical tuning of graphene-loaded plasmonic antennas," *Nano Lett.*, vol. 13, no. 3, pp. 1257–1264, 2013.

[146] Y. Huang, L.-S. Wu, M. Tang, and J. Mao, "Design of a Beam Reconfigurable THz Antenna With Graphene-Based Switchable High-Impedance Surface," *Nanotechnology, IEEE Trans.*, vol. 11, no. 4, pp. 836–842, 2012.

[147] Z. Xu, X. Dong, and J. Bornemann, "Design of a reconfigurable MIMO system for THz communications based on graphene antennas," *IEEE Trans. Terahertz Sci. Technol.*, vol. 4, no. 5, pp. 609–617, 2014.

[148] D. R. Jackson and A. A. Oliner, "Leaky-wave antennas," *Mod. Antenna Handb.*, pp. 325–367, 2008.

[149] M. Guglielmi and D. R. Jackson, "Broadside Radiation from Periodic Leaky-Wave Antennas," *IEEE Trans. Antennas Propag.*, vol. 41, no. 1, pp. 31–37, 1993.

[150] F. Mesa, D. R. Jackson, and M. J. Freire, "Evolution of leaky modes on printed-circuit lines," *IEEE Trans. Microw. Theory Tech.*, vol. 50, no. 1 I, pp. 94–104, 2002.

[151] D. R. Jackson, A. A. Oliner, and A. Ip, "Leaky-wave propagation and radiation for a narrow-beam multiple-layer dielectric structure," *IEEE Trans. Antennas Propag.*, vol. 41, no. 3, pp. 344–348, 1993.

[152] C. Caloz and T. Itoh, *Electromagnetic Metamaterials: Transmission Line Theory and Microwave Applications*. John Wiley & Sons, 2005.

[153] G. V. Eleftheriades and K. G. Balmain, *Negative-Refraction Metamaterials: Fundamental Principles and Applications*. John Wiley & Sons, 2005.

[154] J. S. Gomez-Diaz, A. Álvarez-Melcon, and T. Bertuch, "A modal-based iterative circuit model for the analysis of CRLH leaky-wave antennas comprising periodically loaded PPW," *IEEE Trans. Antennas Propag.*, vol. 59, no. 4, pp. 1101–1112, Apr. 2011.

[155] J. S. Gomez-Diaz, D. Cañete-Rebenaque, and A. Alvarez-Melcon, "A simple CRLH LWA circuit condition for constant radiation rate," *IEEE Antennas*

*Wirel. Propag. Lett.*, vol. 10, pp. 29–32, Mar. 2011.

[156] F. Monticone and A. Alù, "Leaky-Wave Theory , Techniques , and Applications : From Microwaves to Visible Frequencies," *Proc. IEEE*, vol. 103, no. 5, pp. 793–821, 2015.

[157] R. Guzmán-Quirós, J. L. Gómez-Tornero, A. R. Weily, and Y. J. Guo, "Electronically steerable 1-d fabry-perot leaky-wave antenna employing a tunable high impedance surface," *IEEE Trans. Antennas Propag.*, vol. 60, no. 11, pp. 5046–5055, 2012.

[158] J. L. Gómez-Tornero, A. de la Torre Martínez, D. C. Rebenaque, M. Gugliemi, and A. Álvarez-Melcón, "Design of tapered leaky-wave antennas in hybrid waveguide-planar technology for millimeter waveband applications," *IEEE Trans. Antennas Propag.*, vol. 53, no. 8 I, pp. 2563–2577, 2005.

[159] P. Burghignoli, G. Lovat, F. Capolino, D. R. Jackson, and D. R. Wilton, "Directive leaky-wave radiation from a dipole source in a wire-medium slab," *IEEE Trans. Antennas Propag.*, vol. 56, no. 5, pp. 1329–1339, 2008.

[160] Y. Hu *et al.*, "Extraordinary Optical Transmission in Metallic Nanostructures with a Plasmonic Nanohole Array of Two Connected Slot Antennas," *Plasmonics*, pp. 1–6, Dec. 2014.

[161] S. Kim, M. S. Jang, V. W. Brar, Y. Tolstova, K. W. Mauser, and H. A. Atwater, "Electronically tunable extraordinary optical transmission in graphene plasmonic ribbons coupled to subwavelength metallic slit arrays," *Nat. Commun.*, vol. 7, p. 12323, 2016.

[162] Y. Hadad, D. L. Sounas, and A. Alu, "Space-time gradient metasurfaces," *Phys. Rev. B - Condens. Matter Mater. Phys.*, vol. 92, no. 10, 2015.

[163] A. A. Oliner, "Leaky waves: Basic properties and applications," in *Microwave Conference Proceedings, 1997. APMC'97, 1997 Asia-Pacific*, 1997, vol. 1, pp. 397–400.

[164] A. A. Oliner and A. Hessel, "Guided Waves on Sinusoidally-Modulated Reactance Surfaces," *IRE Trans. Antennas Propag.*, vol. 7, no. 5, pp. 201–208, 1959.

[165] G. Minatti, F. Caminita, M. Casaletti, and S. Maci, "Spiral leaky-wave antennas based on modulated surface impedance," *IEEE Trans. Antennas Propag.*, vol. 59, no. 12, pp. 4436–4444, 2011.

[166] A. M. Patel and A. Grbic, "A printed leaky-wave antenna based on a sinusoidally-modulated reactance surface," *IEEE Trans. Antennas Propag.*, vol. 59, no. 6 PART 2, pp. 2087–2096, 2011.

[167] A. Grbic and G. V. Eleftheriades, "Leaky CPW-Based Slot Antenna Arrays for Millimeter-Wave Applications," *IEEE Trans. Antennas Propag.*, vol. 50, no. 11, pp. 1494–1504, Nov. 2002.

[168] M. Ettorre and A. Grbic, "Generation of Propagating Bessel Beams Using Leaky-Wave Modes," *IEEE Trans. Antennas Propag.*, vol. 60, no. 8, pp. 3605–3613, 2012.

[169] X. C. Wang, W. S. Zhao, J. Hu, and W. Y. Yin, "Reconfigurable terahertz leaky-wave antenna using graphene-based high-impedance surface," *IEEE*

[170] W. Fuscaldo, P. Burghignoli, P. Baccarelli, and A. Galli, "A Reconfigurable Substrate–Superstrate Graphene-Based Leaky-Wave THz Antenna," *IEEE Antennas Wirel. Propag. Lett.*, vol. 15, pp. 1545–1548, 2016.

[171] W. Fuscaldo, P. Burghignoli, P. Baccarelli, and A. Galli, "A graphene-loaded substrate-superstrate leaky-wave THz antenna," in *2016 10th European Conference on Antennas and Propagation (EuCAP)*, 2016, pp. 1–3.

[172] F. Ruesink, M.-A. Miri, A. Alù, and E. Verhagen, "Nonreciprocity and magnetic-free isolation based on optomechanical interactions," *Nat. Commun.*, pp. 1–16, 2016.

[173] S. Taravati and C. Caloz, "Mixer-Duplexer-Antenna Leaky-Wave System Based on Periodic Space-Time Modulation," *IEEE Trans. Antennas Propag.*, vol. 65, no. 2, pp. 442–452, Feb. 2017.

[174] D. Correas-Serrano, J. S. Gomez-Diaz, D. L. Sounas, A. Alvarez-Melcon, and A. Alù, "Non-reciprocal THz components based on spatiotemporally modulated graphene," in *2016 10th European Conference on Antennas and Propagation (EuCAP)*, 2016, pp. 1–4.

[175] J. Huang and J. A. Encinar, *Reflectarray Antennas*. Hoboken, NJ, USA: John Wiley & Sons, Inc., 2007.

[176] J. A. Encinar, "Design of two-layer printed reflectarrays using patches of variable size," *IEEE Trans. Antennas Propag.*, vol. 49, no. 10, pp. 1403–1410, 2001.

[177] P. Dreyer, J. S. Gomez Diaz, and J. Perruisseau-Carrier, "Design of a reflectarray element integrated in a Solar Cell panel," *IEEE Antennas Propag. Soc. AP-S Int. Symp.*, pp. 1558–1559, 2013.

[178] J. M. Baracco, P. Ratajczak, P. Brachat, and G. Toso, "Dual frequency Ka-band reflectarray for ground terminal application," *8th Eur. Conf. Antennas Propagation, EuCAP 2014*, pp. 1437–1440, 2014.

[179] J. A. Encinar, M. Arrebola, F. Luis, and G. Toso, "A transmit-receive reflectarray antenna for direct broadcast satellite applications," *IEEE Trans. Antennas Propag.*, vol. 59, no. 9, pp. 3255–3264, 2011.

[180] H. Legay, B. Pinte, M. Charrier, A. Ziaei, E. Girard, and R. Gillard, "A steerable reflectarray antenna with MEMS controls," in *Phased Array Systems and Technology, 2003. IEEE International Symposium on*, 2003, pp. 494–499.

[181] F. Monticone, N. M. Estakhri, and A. Alu, "Full control of nanoscale optical transmission with a composite metascreen," *Phys. Rev. Lett.*, vol. 110, no. 20, pp. 1–5, May 2013.

[182] N. Yu *et al.*, "Light propagation with phase discontinuities: generalized laws of reflection and refraction.," *Science (80-. ).*, vol. 334, no. October, pp. 333–337, Oct. 2011.

[183] A. V Kildishev, A. Boltasseva, and V. M. Shalaev, "Planar Photonics with Metasurfaces," *Science (80-. ).*, vol. 339, no. 6125, p. 1232009, 2013.

[184] S. Sun, Q. He, S. Xiao, Q. Xu, X. Li, and L. Zhou,

"Gradient-index meta-surfaces as a bridge linking propagating waves and surface waves," *Nat. Mater.*, vol. 11, no. 5, pp. 426–431, 2012.

[185] N. Mohammadi Estakhri and A. Alù, "Wave-front Transformation with Gradient Metasurfaces," *Phys. Rev. X*, vol. 6, no. 4, p. 41008, 2016.

[186] X. Ni, N. K. Emani, A. V. Kildishev, A. Boltasseva, and V. M. Shalaev, "Broadband Light Bending with Plasmonic Nanoantennas," *Science (80-. ).*, vol. 335, no. 6067, pp. 427–427, 2012.

[187] S. V Hum, M. Okoniewski, and R. J. Davies, "Realizing an electronically tunable reflectarray using varactor diode-tuned elements," *IEEE Microw. Wirel. Components Lett.*, vol. 15, no. 6, pp. 422–424, 2005.

[188] L. Boccia, F. Venneri, G. Amendola, and G. Di Massa, "Application of varactor diodes for reflectarray phase control," in *IEEE Antennas and Propagation Society International Symposium*, 2002, vol. 3, pp. 132–135.

[189] E. Carrasco, M. Barba, and J. A. Encinar, "Reflectarray element based on aperture-coupled patches with slots and lines of variable length," *IEEE Trans. Antennas Propag.*, vol. 55, no. 3, pp. 820–825, 2007.

[190] J. Perruisseau-Carrier and A. K. Skrivervik, "Monolithic MEMS-based reflectarray cell digitally reconfigurable over a 360 phase range," *IEEE Antennas Wirel. Propag. Lett.*, vol. 7, pp. 138–141, 2008.

[191] S. V Hum, G. McFeetors, and M. Okoniewski, "Integrated MEMS reflectarray elements," in *2006 First European Conference on Antennas and Propagation*, 2006, pp. 1–6.

[192] W. Hu *et al.*, "Liquid-crystal-based reflectarray antenna with electronically switchable monopulse patterns," *Electron. Lett.*, vol. 43, no. 14, p. 1, 2007.

[193] A. Moessinger, R. Marin, S. Mueller, J. Freese, and R. Jakoby, "Electronically reconfigurable reflectarrays with nematic liquid crystals," *Electron. Lett.*, vol. 42, no. 16, p. 1, 2006.

[194] J. Verbeeck, H. Tian, and P. Schattschneider, "Production and application of electron vortex beams.," *Nature*, vol. 467, no. 7313, pp. 301–304, 2010.

[195] B. J. McMorran *et al.*, "Electron vortex beams with high quanta of orbital angular momentum.," *Science*, vol. 331, no. 6014, pp. 192–195, 2011.

[196] Q. Zhan, "Properties of circularly polarized vortex beams," *Opt. Lett.*, vol. 31, no. 7, pp. 867–869, 2006.

[197] G. V. Bogatyryova, C. V. Fel'de, P. V. Polyanskii, S. A. Ponomarenko, M. S. Soskin, and E. Wolf, "Partially coherent vortex beams with a separable phase," *Opt. Lett.*, vol. 28, no. 11, p. 878, 2003.

[198] K. Y. Bliokh, I. V Shadrivov, and Y. S. Kivshar, "Goos-Hänchen and Imbert-Fedorov shifts of polarized vortex beams.," *Opt. Lett.*, vol. 34, no. 3, pp. 389–391, 2009.

[199] G. Li *et al.*, "Spin-enabled plasmonic metasurfaces for manipulating orbital angular momentum of light," *Nano Lett.*, vol. 13, no. 9, pp. 4148–4151, 2013.

[200] M. Q. Mehmood *et al.*, "Visible-Frequency Metasurface for Structuring and Spatially Multiplexing Optical Vortices," *Adv. Mater.*, p. n/a-n/a, Jan. 2016.

[201] F. Tamburini, E. Mari, A. Sponselli, A. Bianchini, and F. Romanato, "Encoding many channels on the same frequency through radio vorticity and oam techniques," *Semin. Technol. Bachelor O F Eng. Commun.*, vol. 14, no. 3, p. 33001, 2012.

[202] M. Tamagnone, C. Craeye, and J. Perruisseau-Carrier, "Comment on 'Encoding many channels on the same frequency through radio vorticity: first experimental test,'" *New J. Phys.*, vol. 14, no. 11, p. 118001, 2012.

[203] F. Tamburini, B. Thidé, E. Mari, A. Sponselli, A. Bianchini, and F. Romanato, "Reply to comment on 'Encoding many channels on the same frequency through radio vorticity: First experimental test,'" *New J. Phys.*, vol. 14, 2012.

[204] W. Cheng, "Optical vortex beams: Generation, propagation and applications," PhD Dissertation, University of Dayton, 2013.

[205] A. S. Desyatnikov, Y. S. Kivshar, and L. Torner, "Optical vortices and vortex solitons," *Prog. Opt.*, vol. 47, pp. 291–391, 2005.

[206] Z. Chang *et al.*, "A Reconfigurable Graphene Reflectarray for Generation of Vortex THz Waves," *IEEE Antennas Wirel. Propag. Lett.*, vol. 15, pp. 1537–1540, 2016.

[207] J. Wong, M. Selvanayagam, and G. V. Eleftheriades, "A thin printed metasurface for microwave refraction," in *International Microwave Symposium (IMS)*, 2014, pp. 1–4.

[208] G. W. Hanson, "Quasi-transverse electromagnetic modes supported by a graphene parallel-plate waveguide," *J. Appl. Phys.*, vol. 104, no. 8, p. 84314, 2008.

[209] D. Correas-Serrano, J. S. Gomez-Diaz, J. Perruisseau-Carrier, and A. Alvarez-Melcon, "Study of spatial dispersion in graphene parallel-plate waveguides and equivalent circuit," *8th Eur. Conf. Antennas Propagation, EuCAP 2014*, pp. 2674–2677, 2014.

[210] F. Rana, "Graphene Terahertz Plasmon Oscillators," *IEEE Trans. Nanotechnol.*, vol. 7, no. 1, pp. 91–99, Jan. 2008.

[211] D. M. Pozar, *Microwave engineering*. John Wiley & Sons, 2009.

[212] C. A. Balanis, *Advanced Engineering electromagnetics*. John Wiley and Sons, 2012.

[213] J. S. Gómez-Díaz and J. Perruisseau-Carrier, "Graphene-based plasmonic switches at near infrared frequencies," *Opt. Express*, vol. 21, no. 13, p. 15490, Feb. 2013.

[214] P.-Y. Chen, H. Huang, D. Akinwande, and A. Alu, "Graphene-Based Plasmonic Platform for Reconfigurable Terahertz Nanodevices," *ACS Photonics*, vol. 1, no. 8, pp. 647–654, Aug. 2014.

[215] P.-Y. Chen, C. Argyropoulos, and A. Alu, "Terahertz antenna phase shifters using integrally-gated graphene transmission-lines," *IEEE Trans. Antennas Propag.*, vol. 61, no. 4, pp. 1528–1537, 2013.

[216] K. M. Milaninia, M. A. Baldo, A. Reina, and J. Kong, "All graphene electromechanical switch fabricated by chemical vapor deposition," *Appl. Phys. Lett.*, vol. 183105, no. 2009, pp. 1–4, 2009.

[217] R. J. Cameron *et al.*, *Microwave Filters for Communication Systems: Fundamentals, Design and Applications*. Wiley-Blackwell, 2007.

[218] P. V. Castejon, D. C. Serrano, F. D. Q. Pereira, J. Hinojosa, and A. Á. Melcon, "A novel low-pass filter based on rounded posts designed by an alternative full-wave analysis technique," *IEEE Trans. Microw. Theory Tech.*, vol. 62, no. 10, pp. 2300–2307, 2014.

[219] A. E. Atia and A. E. Williams, "Narrow-Bandpass Waveguide Filters," *IEEE Trans. Microw. Theory Tech.*, vol. 20, no. 4, pp. 258–265, Apr. 1972.

[220] R. Levy and S. B. Cohn, "A History of Microwave Filter Research, Design, and Development," *IEEE Trans. Microw. Theory Tech.*, vol. 32, no. 9, pp. 1055–1067, Sep. 1984.

[221] L. Novotny and B. Hecht, *Principles of nano-optics*. Cambridge university press, 2012.

[222] L. Novotny, "Effective wavelength scaling for optical antennas," *Phys. Rev. Lett.*, vol. 98, no. 26, p. 266802, 2007.

[223] P. Mühlschlegel, H.-J. Eisler, O. J. F. Martin, B. Hecht, and D. W. Pohl, "Resonant optical antennas," *Science (80-. ).*, vol. 308, no. 5728, pp. 1607–1609, 2005.

[224] D. Correas-Serrano, J. S. Gomez-Diaz, and A. Alvarez-Melcon, "On the influence of spatial dispersion on the performance of graphene-based plasmonic devices," *IEEE Antennas Wirel. Propag. Lett.*, vol. 13, pp. 345–348, 2014.

[225] S. Raza, S. I. Bozhevolnyi, M. Wubs, and N. A. Mortensen, "Nonlocal optical response in metallic nanostructures," *J. Phys. Condens. Matter*, vol. 27, no. 4, p. 183204, Jun. 2015.

[226] G. Toscano *et al.*, "Nonlocal response in plasmonic waveguiding with extreme light confinement," *Nanophotonics*, vol. 2, no. 3, pp. 161–166, 2013.

[227] T. Christensen, W. Yan, S. Raza, A. P. Jauho, N. A. Mortensen, and M. Wubs, "Nonlocal response of metallic nanospheres probed by light, electrons, and atoms," *ACS Nano*, vol. 8, no. 2, pp. 1745–1758, 2014.

[228] W. Yan, M. Wubs, and N. A. Mortensen, "Hyperbolic metamaterials: Nonlocal response regularizes broadband supersingularity," *Phys. Rev. B - Condens. Matter Mater. Phys.*, vol. 86, no. 20, pp. 1–8, 2012.

[229] S. Raza, G. Toscano, A. P. Jauho, M. Wubs, and N. A. Mortensen, "Unusual resonances in nanoplasmonic structures due to nonlocal response," *Phys. Rev. B - Condens. Matter Mater. Phys.*, vol. 84, no. 12, pp. 1–5, 2011.

[230] A. Fallahi, T. Low, M. Tamagnone, and J. Perruisseau-Carrier, "Nonlocal electromagnetic response of graphene nanostructures," *Phys. Rev. B - Condens. Matter Mater. Phys.*, vol. 91, no. 12, pp. 1–5, Oct. 2015.

[231] J. S. Gomez-Diaz, M. Tymchenko, and A. Alu, "Hyperbolic Plasmons and Topological Transitions over Uniaxial Metasurfaces," *Phys. Rev. Lett.*, vol. 114, no. 23, p. 233901, 2015.

[232] J. S. Gomez-Diaz and A. Alu, "Flatland Optics with Hyperbolic Metasurfaces," *ACS Photonics*, vol. 3, no. 12, p. acsphotonics.6b00645, 2016.

[233] O. Takayama, L.-C. Crasovan, S. K. Johansen, D. Mihalache, D. Artigas, and L. Torner, "Dyakonov Surface Waves: A Review," *Electromagnetics*, vol. 28, no. 3, pp. 126–145, 2008.

[234] O. Takayama, L. Crasovan, D. Artigas, and L. Torner, "Observation of dyakonov surface waves," *Phys. Rev. Lett.*, vol. 102, no. 4, pp. 2–5, 2009.

[235] A. a. High *et al.*, "Visible-frequency hyperbolic metasurface," *Nature*, vol. 522, no. 7555, pp. 192–196, 2015.

[236] M. Ghorbanzadeh, S. Darbari, and M. K. Moravvej-Farshi, "Graphene-based plasmonic force switch," *Appl. Phys. Lett.*, vol. 108, no. 11, pp. 2–4, 2016.

[237] M. Dragoman, D. Dragoman, F. Coccetti, R. Plana, and A. A. Muller, "Microwave switches based on graphene," *J. Appl. Phys.*, vol. 105, no. 5, pp. 1–4, 2009.

[238] F. Chen, D. Yao, and Y. Liu, "Graphene-metal hybrid plasmonic switch," *Appl. Phys. Express*, vol. 7, no. 8, 2014.

[239] N. Chamanara, D. Sounas, and C. Caloz, "Non-reciprocal magnetoplasmon graphene coupler," *Opt. Express*, vol. 21, no. 9, pp. 648–652, 2013.

[240] M. Shalaby, M. Peccianti, Y. Ozturk, and R. Morandotti, "A magnetic non-reciprocal isolator for broadband terahertz operation," *Nat. Commun.*, vol. 4, p. 1558, 2013.

[241] M. Tamagnone *et al.*, "Near optimal graphene terahertz non-reciprocal isolator," *Nat. Commun.*, vol. 7, pp. 1–6, 2016.

[242] J. M. Jornet and I. F. Akyildiz, "Graphene-based plasmonic nano-transceiver for terahertz band communication," *8th Eur. Conf. Antennas Propag.*, pp. 492–496, 2014.

[243] A. Menikh, R. MacColl, C. A. Mannella, and X. C. Zhang, "Terahertz biosensing technology: Frontiers and progress," *ChemPhysChem*, vol. 3, no. 8, pp. 655–658, 2002.

[244] C. Debus and P. H. Bolivar, "Terahertz biosensors based on double split ring arrays," *Metamaterials*, vol. 6987, no. November, 2008.

[245] B. R. Brucherseifer M., Nagel M, Haring Bolivar P, Kurz H, Bosserhoff A, "Label-free probing of the binding state of DNA by time-domain terahertz sensing," *Appl. Phys. Lett.*, vol. 77, no. 24, pp. 4049–4051, 2000.

[246] A. . G. Markelz, A. Roitberg, and E. . J. Heilweil, "Pulsed terahertz spectroscopy of DNA, bovine serum albumin and collagen between 0.1 and 2.0 THz," *Chem. Phys. Lett.*, vol. 320, no. March, pp. 42–48, 2000.

[247] B. S. Alexandrov, V. Gelev, A. R. Bishop, A. Usheva, and K. Rasmussen, "DNA breathing dynamics in the presence of a terahertz field," *Phys. Lett. Sect. A Gen.*

*At. Solid State Phys.*, vol. 374, no. 10, pp. 1214–1217, 2010.

[248] B. M. Fischer, M. Walther, and P. Uhd Jepsen, "Far-infrared vibrational modes of DNA components studied by terahertz time-domain spectroscopy.," *Phys. Med. Biol.*, vol. 47, no. 21, pp. 3807–3814, 2002.

[249] D. Rodrigo *et al.*, "Mid-infrared plasmonic biosensing with graphene," *Science (80-. ).*, vol. 349, no. 6244, pp. 165–168, 2015.

[250] J. N. Anker, W. P. Hall, O. Lyandres, N. C. Shah, J. Zhao, and R. P. Van Duyne, "Biosensing with plasmonic nanosensors.," *Nat. Mater.*, vol. 7, no. 6, pp. 442–453, 2008.

[251] J. Homola *et al.*, "Surface plasmon resonance sensors: review," *Sensors Actuators B Chem.*, vol. 54, no. 1–2, pp. 3–15, Jan. 1999.

[252] L. Tong, H. Wei, S. Zhang, and H. Xu, "Recent advances in plasmonic sensors," *Sensors (Basel).*, vol. 14, no. 5, pp. 7959–7973, Jan. 2014.

[253] F. Huang *et al.*, "Zero-Reflectance Metafilms for Optimal Plasmonic Sensing," *Adv. Opt. Mater.*, vol. 4, no. 2, pp. 328–335, Nov. 2016.

[254] A. G. Brolo, R. Gordon, B. Leathem, and K. L. Kavanagh, "Surface Plasmon Sensor Based on the Enhanced Light Transmission through Arrays of Nanoholes in Gold Films," *Langmuir*, no. 17, pp. 4813–4815, 2004.

[255] O. V Shapoval and A. I. Nosich, "Bulk refractive-index sensitivities of the THz-range plasmon resonances on a micro-size graphene strip," *J. Phys. D. Appl. Phys.*, vol. 49, no. 5, p. 055105 1-9, 2016.

[256] T. Zinenko, A. Matsushima, and A. Nosich, "Surface Plasmon, Grating-Mode and Slab-Mode Resonances in the H- and E-polarized THz Wave Scattering by a Graphene Strip Grating Embedded into a Dielectric Slab," *IEEE J. Sel. Top. Quantum Electron.*, vol. 23, no. 6, pp. 1–1, 2017.

[257] H. Huang, M. Sakhdari, M. Hajizadegan, A. Shahini, D. Akinwande, and P.-Y. Chen, "Toward transparent and self-activated graphene harmonic transponder sensors," *Appl. Phys. Lett.*, vol. 108, no. 17, 2016.

[258] C. Argyropoulos, "Enhanced transmission modulation based on dielectric metasurfaces loaded with graphene," *Opt. Express*, vol. 23, no. 18, p. 23787, 2015.

[259] J. S. Gomez-Diaz, M. Tymchenko, and A. Alù, "Hyperbolic metasurfaces: surface plasmons, light-matter interactions, and physical implementation using graphene strips," *Opt. Mater. Express*, vol. 5, no. 10, p. 2313, Oct. 2015.

[260] Y. Liu and X. Zhang, "Metasurfaces for manipulating surface plasmons," *Appl. Phys. Lett.*, vol. 103, no. 14, 2013.

[261] S. A. H. Gangaraj, T. Low, A. Nemilentsau, and G. Hanson, "Directive Surface Plasmons on Tunable Two-Dimensional Hyperbolic Metasurfaces and Black Phosphorus: Green's Function and Complex Plane Analysis," *IEEE Trans. Antennas Propag.*, vol. PP, no. 99, p. 1, 2016.

[262] A. Nemilentsau, T. Low, and G. Hanson, "Anisotropic 2D materials for tunable hyperbolic plasmonics," *Phys. Rev. Lett.*, vol. 116, no. 6, pp. 1–5, Feb. 2016.

[263] M. a K. Othman, C. Guclu, and F. Capolino, "Graphene-based tunable hyperbolic metamaterials and enhanced near-field absorption.," *Opt. Express*, vol. 21, no. 6, pp. 7614–32, Mar. 2013.

[264] G. Tianjing and C. Argyropoulos, "Broadband polarizers based on graphene metasurfaces," *Opt. Lett.*, vol. 41, no. 23, pp. 5592–5595, 2016.

[265] T. Christensen, W. Yan, A. P. Jauho, M. Wubs, and N. A. Mortensen, "Kerr nonlinearity and plasmonic bistability in graphene nanoribbons," *Phys. Rev. B - Condens. Matter Mater. Phys.*, vol. 92, no. 12, pp. 1–10, 2015.

[266] S. A. Mikhailov and K. Ziegler, "Nonlinear electromagnetic response of graphene: frequency multiplication and the self-consistent-field effects.," *J. Phys. Condens. Matter*, vol. 20, no. 38, p. 384204, 2008.

[267] N. Nookala *et al.*, "Ultrathin gradient nonlinear metasurface with a giant nonlinear response," *Optica*, vol. 3, no. 3, p. 283, 2016.

[268] J. Lee *et al.*, "Giant nonlinear response from plasmonic metasurfaces coupled to intersubband transitions.," *Nature*, vol. 511, no. 7507, pp. 65–9, Jul. 2014.

[269] J. Wang, Y. Hernandez, M. Lotya, J. N. Coleman, and W. J. Blau, "Broadband nonlinear optical response of graphene dispersions," *Adv. Mater.*, vol. 21, no. 23, pp. 2430–2435, 2009.

[270] D. A. Smirnova, A. E. Miroshnichenko, Y. S. Kivshar, and A. B. Khanikaev, "Tunable nonlinear graphene metasurfaces," *Phys. Rev. B - Condens. Matter Mater. Phys.*, vol. 92, no. 16, pp. 1–6, 2015.

[271] C. Argyropoulos, P. Y. Chen, F. Monticone, G. D'Aguanno, and A. Alu, "Nonlinear plasmonic cloaks to realize giant all-optical scattering switching," *Phys. Rev. Lett.*, vol. 108, no. 26, p. 263905, Jun. 2012.

[272] C. Argyropoulos, N. M. Estakhri, F. Monticone, and A. Alu, "Negative refraction, gain and nonlinear effects in hyperbolic metamaterials," *Opt. Express*, vol. 21, no. 12, pp. 15037–15047, 2013.

[273] J. S. Gomez-Diaz, M. Tymchenko, J. Lee, M. A. Belkin, and A. Alù, "Nonlinear processes in multi-quantum-well plasmonic metasurfaces: Electromagnetic response, saturation effects, limits, and potentials," *Phys. Rev. B - Condens. Matter Mater. Phys.*, vol. 92, no. 12, pp. 1–13, 2015.

[274] Z. Qian, Y. Hui, F. Liu, S. Kang, S. Kar, and M. Rinaldi, "Graphene-aluminum nitride NEMS resonant infrared detector," *Microsystems Nanoeng.*, vol. 2, 2016.

[275] Y. Hui, J. S. Gomez-Diaz, Z. Qian, A. Alù, and M. Rinaldi, "Plasmonic piezoelectric nanomechanical resonator for spectrally selective infrared sensing," *Nat. Commun.*, vol. 7, p. 11249, 2016.

[276] T. Otsuji, "Trends in the Research of Modern Terahertz Detectors: Plasmon Detectors," *IEEE*

*Trans. Terahertz Sci. Technol.*, vol. 5, no. 6, pp. 1110–1120, Nov. 2015.

[277] W. Knap *et al.*, "Field effect transistors for terahertz detection: Physics and first imaging applications," *J. Infrared, Millimeter, Terahertz Waves*, vol. 30, no. 12, pp. 1319–1337, 2009.

[278] D. Mittleman, *Sensing with terahertz radiation*. Springer, 2013.

[279] X. G. Peralta *et al.*, "Terahertz photoconductivity and plasmon modes in double-quantum-well field-effect transistors," *Appl. Phys. Lett.*, vol. 81, no. 9, pp. 1627–1629, 2002.

[280] W. Knap *et al.*, "Resonant detection of subterahertz radiation by plasma waves in a submicron field-effect transistor," *Appl. Phys. Lett.*, vol. 80, no. 18, pp. 3433–3435, 2002.

[281] W. Knap *et al.*, "Nonresonant detection of terahertz radiation in field effect transistors," *J. Appl. Phys.*, vol. 91, no. 11, pp. 9346–9353, 2002.

[282] W. Knap *et al.*, "Plasma wave detection of sub-terahertz and terahertz radiation by silicon field-effect transistors," *Appl. Phys. Lett.*, vol. 85, no. 4, pp. 675–677, 2004.

[283] F. Teppe *et al.*, "Room-temperature plasma waves resonant detection of sub-terahertz radiation by nanometer field-effect transistor," *Appl. Phys. Lett.*, vol. 87, no. 5, pp. 2005–2007, 2005.

[284] B. B. Hu and M. C. Nuss, "Imaging with terahertz waves," *Opt. Lett.*, vol. 20, no. 16, pp. 1716–1718, 1995.

[285] D. M. Mittleman, R. H. Jacobsen, and M. C. Nuss, "T-ray imaging," *IEEE J. Sel. Top. quantum Electron.*, vol. 2, no. 3, pp. 679–692, 1996.

[286] T. Löffler, T. Bauer, K. J. Siebert, H. G. Roskos, A. Fitzgerald, and S. Czasch, "Terahertz dark-field imaging of biomedical tissue," *Opt. Express*, vol. 9, no. 12, pp. 616–621, 2001.

[287] M. Mittendorff *et al.*, "Ultrafast graphene-based broadband THz detector," *Appl. Phys. Lett.*, vol. 103, no. 2, 2013.

[288] T. Otsuji, A. Satou, S. B. Tombet, V. Ryzhii, V. V. Popov, and M. S. Shur, "Graphene plasmonic heterostructures for terahertz device applications," in *Proceedings - 2014 International Conference Laser Optics, LO 2014*, 2014, pp. 1–1.

[289] T. Mueller, F. Xia, and P. Avouris, "Graphene photodetectors for high-speed optical communications," *Nat. Photonics*, vol. 4, no. 5, pp. 297–301, 2010.

[290] V. Ryzhii and M. Ryzhii, "Graphene bilayer field-effect phototransistor for terahertz and infrared detection," *Phys. Rev. B - Condens. Matter Mater. Phys.*, vol. 79, no. 24, pp. 1–8, 2009.

[291] X. Cai *et al.*, "Sensitive room-temperature terahertz detection via the photothermoelectric effect in graphene," *Nat. Nanotechnol.*, vol. 9, no. 10, pp. 814–819, 2014.

[292] G. Liang *et al.*, "Integrated Terahertz Graphene Modulator with 100% Modulation Depth," *ACS Photonics*, vol. 2, no. 11, pp. 1559–1566, 2015.

[293] B. Sensale-Rodriguez *et al.*, "Extraordinary control of terahertz beam reflectance in graphene electro-absorption modulators," *Nano Lett.*, vol. 12, no. 9, pp. 4518–4522, 2012.

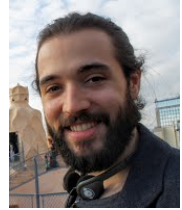

**Diego Correas-Serrano** was born in Jaen, Spain. He received the Telecommunications Engineer degree and the Master on Electrical and Computer Engineering from the Technical University of Cartagena (UPCT), Spain, in 2013 and 2015, both with honors. He is currently a graduate research assistant at the University of California, Davis.

In 2012 he joined the Information Technologies group at UPCT, where he proposed novel waveguide filter topologies for space applications with power handling capabilities surpassing the state of the art. Throughout his master and PhD his interests shifted to plasmonics and metasurfaces using 2D materials, with special attention to reconfigurability and nonreciprocity through spatiotemporal modulation.

In 2015 and 2016 he held visiting scholar positions at the Metamaterial and Plasmonic Research Laboratory of the University of Texas at Austin. He received the 2016 IEEE Antennas and Propagation Society Doctoral Research Grant, the 2014 IEEE Antennas and Propagation Society Pre-Doctoral Research Grant, the 2014 TICRA Foundation Young Scientist Grant at the European Conference on Antennas and Propagation, the 2014 Spain's Ministry of Education FPU Fellowship for doctoral studies. He serves as a reviewer for several microwaves, optics, and physics journals.

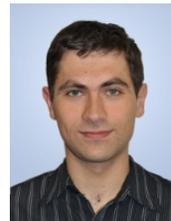

**J. Sebastian Gomez-Diaz** was born in Ontur, Albacete, Spain. He is an Assistant Professor in the Department of Electrical and Computer Engineering at the University of California, Davis.

He received the Telecommunications Engineer and the PhD. degrees (both with honors) from the Technical University of Cartagena (UPCT), Spain, in 2006 and 2011, respectively. During his PhD. he held visiting research positions at the ETA group of Polygrames, École Polytechnique de Montréal (Canada, 1 year) and at the Fraunhofer Institute for High Frequency Physics and Radar Techniques (Germany, 6 months). From October 2011 until March 2014, Dr. Gomez-Diaz was a postdoctoral fellow at the École Polytechnque Fédéral de Lausanne (Switzerland). Then, from May 2014 to August 2016, he continued his postdoctoral work in the Metamaterial and Plasmonic Research Laboratory of The University of Texas at Austin (USA). His main research interests include multidisciplinary areas of electromagnetic wave propagation and radiation, metamaterials and metasurfaces, plasmonics, novel 2D materials, antennas, non-linear phenomena, and other emerging topics on applied electromagnetics and nanotechnology.

Dr. Gomez-Diaz was the recipient of the 2017 Leopold B. Felsen Award for Excellence in Electrodynamics given by the European Association of Antennas and Propagation, the Raj Mittra Award presented by the 2015 IEEE Antennas and Propagation Society, the Young Scientist Award of the 2015 URSI Atlantic RadioScience Conference, a FP7 Marie Curie Fellowship from the European Commision in 2012, the Colegio Oficial de Ingenieros de Telecomunicación (COIT/AEIT) award to the best Spanish PhD. thesis in basic information and communication technologies in 2011, and the best PhD. thesis award from the Technical University of Cartagena (2011). He serves as a reviewer for various journals on antennas, microwaves/THz and physics.